\documentclass[12pt]{elsart}
\usepackage{amsmath}
\usepackage{amssymb}
\usepackage{amsfonts}
\usepackage{graphics}
\usepackage{latexsym}
\usepackage{epsfig}

\graphicspath{plots}
\newcommand{\msb}{\overline{MS}}
\newcommand{\ccvc}{c_{\,CVC}}

\newcommand{\gev}{\,\mbox{GeV}}
\newcommand{\mev}{\,\mbox{MeV}}

\newcommand{\msbar}{\overline{MS}}
\newcommand{\psib}{\bar{\psi}}

\newcommand{\asQ}{\left(\frac{\alpha_s(Q^2)}{\pi}\right)}

\newcommand{\eeq}{\end{equation}}
\newcommand{\beqn}{\begin{eqnarray}}
\newcommand{\eeqn}{\end{eqnarray}}
\newcommand{\beqns}{\begin{eqnarray*}}
\newcommand{\eeqns}{\end{eqnarray*}}

\def\lsim{\mathrel{\rlap{\lower4pt\hbox{\hskip1pt$\sim$}}
    \raise1pt\hbox{$<$}}}                % less than or approx. symbol
\def\gsim{\mathrel{\rlap{\lower4pt\hbox{\hskip1pt$\sim$}}
    \raise1pt\hbox{$>$}}}                % greater than or approx. symbol
\begin{document}
\begin{frontmatter}

\vspace*{0.5cm}
\title{\bf Vacuum Polarisation and Hadronic Contribution
 to muon $g-2$ from Lattice QCD}
 \vspace*{-0.9cm}

\author[leip,reg]{M.~G\"ockeler},
\author[edi]{R.~Horsley},
\author[nic,fu]{W.~K\"urzinger},
\author[nic]{D.~Pleiter},
\author[liv]{P.~E.~L.~Rakow},
\author[nic,de]{G.~Schierholz}

\vspace*{0.35cm}

\address[leip]{Institut f\"ur Theoretische Physik, Universit\"at Leipzig,\\
 D-04109 Leipzig, Germany} 
\address[reg]{Institut f\"ur Theoretische Physik, Universit\"at Regensburg,\\
D-93040 Regensburg, Germany}
\address[edi]{School of Physics, University of Edinburgh,\\
Edinburgh EH9 3JZ, UK}
\address[nic]{John von Neumann-Institut f\"ur Computing NIC,\\
Deutsches Elektronen-Synchrotron DESY,\\
D-15735 Zeuthen, Germany} 
\address[fu]{Institut f\"ur Theoretische Physik, Freie Universit\"at Berlin,\\
D-14195 Berlin, Germany}
\address[liv]{Theoretical Physics Division, Department of Mathematical
 Sciences,\\ University of Liverpool, Liverpool L69 3BX, UK} 
\address[de]{Deutsches Elektronen-Synchrotron DESY,\\
D-22603 Hamburg, Germany}

\vspace*{0.5cm}
 \begin{center}
 [QCDSF Collaboration]

{\vspace*{-17.0cm}
 \begin{flushleft}
{\normalsize DESY 03-199} \\
{\normalsize Edinburgh 2003/22} \\
{\normalsize Leipzig LU-ITP 2003/031} \\
{\normalsize Liverpool LTH 615} 
 \end{flushleft}
\vspace*{14.8cm}}

 \end{center} 

 \date{ }      

\vspace*{-0.75cm}

\begin{abstract}
 We compute the vacuum polarisation on the lattice in quenched QCD using
 non-perturbatively improved Wilson fermions.
  Above $Q^2$ of about 2 GeV$^2$ the results are very close
 to the predictions of perturbative QCD. Below this scale we see signs
 of non-perturbative effects which we can describe by the use of dispersion
 relations. We use our results to estimate the light quark contribution 
 to the muon's anomalous magnetic moment. 
 We find the result $ 446(23) \times 10^{-10}$, where the error 
 only includes statistical uncertainties. Finally we make some comments
 on the applicability of the Operator Product Expansion to our data. 

\end{abstract}


\end{frontmatter}

\section{Introduction}

The vacuum polarisation $\Pi(Q^2)$ provides valuable information on the 
interface between perturbative and non-perturbative physics. It has been the 
subject of intensive discussions in the literature.

 The vacuum polarisation tensor is responsible for the
 running of $\alpha_{em}$,   which must be known
 very accurately  for high-precision electro-magnetic calculations. 
 To calculate  the hadronic  contribution to  the anomalous
  magnetic moment of the muon we need to know 
  the vacuum polarisation at scales from $\sim 100$ MeV to
 many GeV. Perturbative QCD will be unreliable 
 at the low end of this scale, so a non-perturbative calculation
 on the lattice would be useful. 

The vacuum polarisation $\Pi(Q^2)$ is defined by
\begin{equation}
\label{vacuum_tensor1}
\Pi_{\mu \nu}(q) = {i} \int \mbox{d}^4 x \; e^{{ i}qx} 
\langle 0|T \, J_\mu (x) J_\nu(0) |0\rangle \equiv 
(q_\mu q_\nu-q^2g_{\mu\nu})\,\Pi(Q^2),  
\end{equation}
where $J_\mu$ is the hadronic electromagnetic current
\begin{equation}
J_{\mu}(x) = \sum_f e_f\,
\bar{\psi}_f (x)\gamma_{\mu} \psi_f(x) = \frac{2}{3} \bar{u}(x)\gamma_\mu u(x)
- \frac{1}{3} \bar{d}(x)\gamma_\mu d(x) + \cdots  
\label{cuur}
\end{equation}
 and $Q^2 \equiv -  q^2 $
 (so that $Q^2 > 0$ for spacelike momenta, $Q^2 < 0$ for timelike). 
 $\Pi$ can be computed on the lattice for spacelike momenta 
 $Q^2 > 0$. 

   $\Pi$ can also be calculated in perturbation theory. 
 $\Pi$ has to be additively renormalised, even the one-loop
 diagram (with no gluons involved) is logarithmically divergent. 
 This renormalisation  implies that the value of $\Pi$ can be
 shifted up and down by a constant 
 depending on scheme and scale without any physical effects. 
 However the $Q^2$ dependence of $\Pi(Q^2)$ is physically
 meaningful, and it
 must be independent of renormalisation scheme or regularisation.

 Experimentally $\Pi$ can be  calculated from data for the 
total cross 
section of $e^+e^- \rightarrow {\rm hadrons}$ with theoretical predictions of 
QCD by means of the dispersion relations~\cite{Reinders}
\begin{equation}
 12 \pi^2 Q^2 \frac{(-1)^n}{n!}\left( \frac{d}{d\, Q^2} \right)^n\, \Pi(Q^2)
 = Q^2\int_{4 m_{\pi}^2}^{\infty} ds \frac{ R(s)}{(s+Q^2)^{n+1}},
 \label{dispersion}
\end{equation}
where
\begin{equation}
R(s)= \frac{\sigma_{e^+e^-\rightarrow {\rm hadrons}}(s)}
{\sigma_{e^+e^-\rightarrow \mu^+\mu^-}(s)}
 = \left(\frac{3 s}{4 \pi \alpha^2_{em} }\right)
 \;\sigma_{e^+e^-\rightarrow {\rm hadrons}}(s) \;.
\label{hhh}
\end{equation}
The first derivative of the vacuum polarisation (the $n=1$ term in
 eq.(\ref{dispersion})) is referred to 
 as  the Adler {$D$-function}~\cite{D_function}:
\begin{equation}
\label{Adler_D}
D(Q^2)=-12\pi^2 Q^2 \,d\Pi(Q^2)/d Q^2.
\end{equation}

   The anomalous magnetic moment of the muon $(g-2)_\mu$
 can be calculated
 to very high order in QED (5 loop), and measured very precisely.
 $(g-2)_\mu$ is more sensitive to high-energy physics than
 $(g-2)_e$, by a factor $m_\mu^2 / m_e^2$, so it is a more
 promising place to look for signs of new physics, but to
 identify new physics we need to know the conventional
 contributions very accurately.
  QED perturbative calculations take good account of
 muon and electron loops, but at the two-loop level quarks
 can be produced, which in turn will produce gluons. The
 dominant contribution comes from
 photons with virtualities $\sim m_\mu^2$, which is a region
 where QCD perturbation theory will not work well.

\begin{figure}[tb] 
\begin{center}
\epsfysize=10cm
\epsfxsize=10cm
\epsfbox{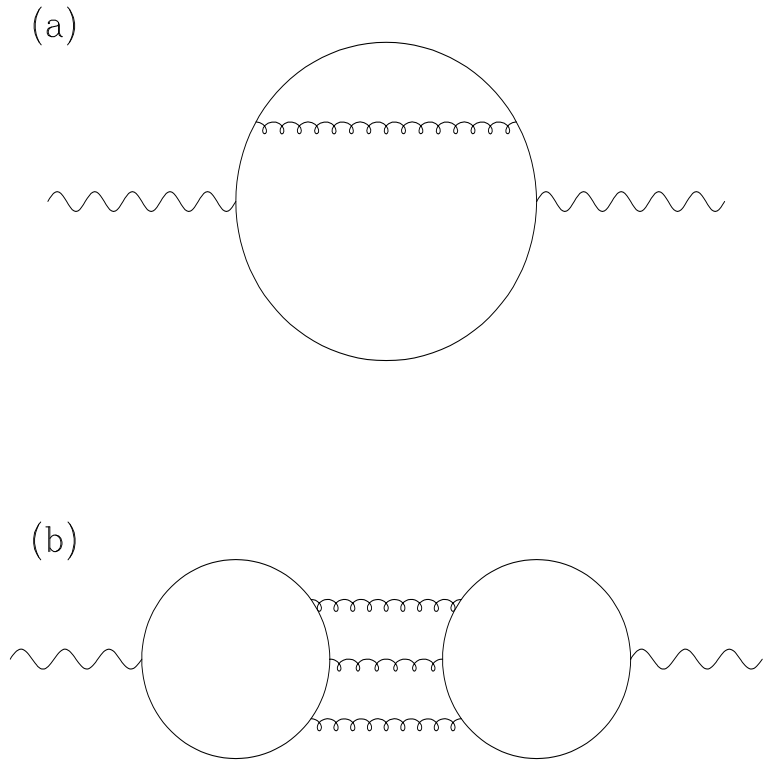}
\vspace*{1.5cm}
\caption{
 A typical Feynman diagram contributing to the $C_\Pi$  
 part of the vacuum polarisation $\Pi$ is shown as diagram (a). 
 A diagram contributing to the $A_\Pi$ part of $\Pi$ 
 is shown in (b). In this paper we only consider diagrams of
 the first type, with both photons attached to the same quark line. 
  \label{feynpol}}
\end{center}
\vspace*{0.5cm}
\end{figure}

  $\Pi(Q^2)$ can be split into two contributions with a different
 dependency on the quark charges, as shown in Fig.\ref{feynpol},
 \begin{equation}
 -12\pi^2 \Pi(Q^2) = \sum_f e_f^2 \; C_\Pi(\mu^2,Q^2,m_f)
 + \sum_{f,f^\prime} e_f e_{f^\prime} \;
 A_\Pi(\mu^2, Q^2, m_f, m_{f^\prime}) . 
 \label{pisplit} 
 \end{equation} 
 The $C_\Pi$ term begins with a tree-level term which is 
 $\mathcal{O}(\alpha_s^0)$, 
 while the first contribution to the $A_\Pi$ term is $\mathcal{O}(\alpha_s^3)$. 
 Furthermore, if flavour $SU(3)$ is a good symmetry, the contribution
 to $A_\Pi$ from the three light flavours ($u, d, s$) cancels because 
 $e_u + e_d +e_s = 0$, and the only surviving contributions to  $A_\Pi$
 come when both $f$ and $f^\prime$ are heavy quarks ($c, b, t$). 
 In this paper we will concentrate on the $C_\Pi$ term, both because it
 is larger, and because it is much easier to measure on the lattice. 

For large (spacelike) momenta $C_\Pi(\mu^2,Q^2,m_f)$ can be expressed
 by means of the Operator Product Expansion, OPE
\begin{eqnarray}
\label{OPE} \lefteqn{
 C_\Pi(\mu^2,Q^2,m_f) =  c_0(\mu^2,Q^2,m_f) 
+ \frac{c_4^F(\mu^2,Q^2)}{(Q^2)^2}\, m_f\, 
\langle \bar{\psi}_f \psi_f \rangle}
  \\[0.3em] &+ \frac{c_4^G(\mu^2,Q^2)}{(Q^2)^2}\, 
\frac{\alpha_s}{\pi}\, \langle G_{\mu\nu}^2\rangle 
 + \frac{{c_4^{\,\prime\,F}}(\mu^2,Q^2)}{(Q^2)^2}\, 
\sum_{f^{\,\prime}} 
m_{f^{\prime}}\, \langle \bar{\psi}_{f^{\prime}} \psi_{f^{\prime}}\rangle
+ \mathcal{O}(1/(Q^2)^3), \nonumber
\end{eqnarray}
where $c_0$, $c_4^F$, $c_4^{\,\prime\,F}$ and $c_4^G$ are the Wilson 
coefficients 
and $\mu$ is the renormalisation scale parameter in some 
 renormalisation scheme such as $\overline{MS}$. The $f'$ sum 
extends over the flavours of the sea quarks
 (internal quark loops not directly connected to the photon lines).
 The Wilson coefficients can be 
computed in perturbation theory, while the non-perturbative physics is 
encoded in the condensates $m_f \, \langle \bar{\psi}_f \psi_f(\mu)\rangle$,
$\alpha_s \langle G_{\mu\nu}^2(\mu)\rangle$, etc. 

   Perturbatively the functions $D$, eq.(\ref{Adler_D}),
 and $R$, eq.(\ref{hhh}), are known to four loops for massless
 quarks~\cite{Gorish,Chetyrkin}, while the coefficient $c_0$ is 
known to three loops 
in the massive case~\cite{Chetyrkin&Kuhn}.  The coefficients $c_4^F$, 
$c_4^{\,\prime\,F}$ and $c_4^G$ are known up to 
$\mathcal{O}(\alpha_s^2)$~\cite{SVZI,SVZII,Wilson_NP} for massless quarks. 

In this paper we shall compute $\Pi(Q^2)$ on the lattice and compare the
result with current phenomenology~\cite{Eidel}. Preliminary results
of this calculation were presented in~\cite{Kuerz}.

  The structure of this paper is as follows. After this Introduction we 
 discuss the lattice setup in Sections~2 and Appendix~A. The
 results are presented in Section~3 and  Appendix~C. In Section~4
 and Appendix~B 
  we compare with perturbation theory.  In Sections~5 and 6
 we present a simple model which describes our lattice data well.
 In Section~7 we use this model to give a lattice estimate of the
 hadronic contribution to the muon's anomalous magnetic moment. 
 Finally in Section~8 we make some comments on the applicability
 of the Operator Product Expansion, and  give our conclusions in
 Section~9. 

\section{Lattice Setup}

We work with non-perturbatively improved Wilson fermions. For the action and
computational details, as well as for results of hadron masses and decay
constants, see~\cite{scaling,dirk,quenchspectro}. 
 Full details of the vacuum polarisation calculation
 can be found in~\cite{Thesis}. 

To discretise
 the hadronic electromagnetic current 
 \begin{equation}
 J_\mu^{em}(x) = \sum_f e_f J_\mu^f(x)
 \end{equation}
 we use the conserved vector current
\begin{equation}
\label{curr1}
 \begin{split}
J_{\mu}^f(x)= &\frac{1}{2}  \big(\bar{\psi}_f (x+a \hat{\mu}) 
(1+\gamma_{\mu}) U_{\mu}^{\dagger}(x) \psi_f (x) 
 \\ &
 -\bar{\psi}_f (x) (1-\gamma_{\mu}) U_{\mu}(x) \psi_f (x+a \hat{\mu}) \big)\, 
   \end{split}
\end{equation}
 where $a$ is the lattice spacing. 
From now on we work with a Euclidean metric and write
\begin{equation}
J_\mu^f(\hat{q})=\sum_x e ^{iq(x+a\hat{\mu}/2)} J^f_\mu(x) 
\end{equation}
with
\begin{equation}
\hat{q}_\mu =\frac{2}{a} \sin \big(\frac{aq_\mu}{2}\big) \, .
\end{equation}

 On the lattice the photon self-energy $\Pi$ is not simply given by
 $\langle J^f_\mu(x) J^f_\nu(0) \rangle$. This is because
 the lattice Feynman rules include vertices
 where any number of photons couple to a quark line  (not just
 the single-photon vertex of the continuum)~\cite{MM}. 
We are therefore led to define
\begin{equation}
\label{vacb}
\Pi_{\mu \nu}(\hat{q}) ={\Pi}_{\mu \nu}^{(1)}(\hat{q})
+{\Pi}_{\mu \nu}^{(2)}(\hat{q})\, ,
\end{equation}
with
\begin{equation}
\label{vacu2b}
\Pi_{\mu \nu}^{(1)}(\hat{q}) = a^4 \sum_{x}  e ^{iq(x+a\hat{\mu}/2
-a\hat{\nu}/2)} \langle J^f_{\mu}(x) J^f_{\nu}(0) \rangle  \;,
\end{equation}
and
\begin{equation}
\label{vacuu1b}
\Pi_{\mu \nu}^{(2)}(\hat{q}) = -a \delta_{\mu\nu}
\langle J_{\mu}^{(2)}(0) \rangle \, .
\end{equation}
 where 
\begin{equation}
\begin{split}
J^{(2)}_{\mu}(x)=&\frac{1}{2}  \big(\bar{\psi} (x+a \hat{\mu})
(1+\gamma_{\mu}) U_{\mu}^{\dagger}(x) \psi (x) \\
&+\bar{\psi} (x) (1-\gamma_{\mu}) U_{\mu}(x) \psi (x+a \hat{\mu})
\big)\, .
\end{split}
\end{equation}
It then follows that (\ref{vacb}) fulfils the Ward identity 
\begin{equation}
\label{Wardb}
\hat{q}_\mu \Pi_{\mu\nu}(\hat{q})
 = \hat{q}_\nu \Pi_{\mu\nu}(\hat{q}) = 0 \, .
\end{equation}

The conserved vector current (\ref{curr1}) is on-shell as well as off-shell
improved~\cite{QCDSFimp} in forward matrix elements. In non-forward
matrix elements, such as the vacuum polarisation, it needs to be further
improved, 
\begin{equation}
J^{\rm imp}_\mu(x) = J_\mu(x) + \frac{1}{2} i a\, \ccvc \partial_\nu
\big(\bar{\psi}(x) \sigma_{\mu\nu} \psi(x)\big)\, ,
\end{equation} 
where we have some freedom in choosing $\partial_\nu$ on the lattice. Any
choice must preserve (\ref{Wardb}), and ideally it should keep
$\mathcal{O}(a^2)$ corrections small. Considering the fact that the conserved
vector current $J_\mu(x)$ `lives' at $x+a\hat{\mu}/2$, the natural 
 (or naive) choice for the improved operator would be
 \begin{eqnarray} 
 \label{naiveJ} 
 J^{\rm imp}_\mu(\hat{q}) &=& J_\mu(\hat{q})
 + \frac{1}{2} i a \ccvc \sum_x e ^{iq(x+a\hat{\mu}/2)}
 \frac{1}{4a} 
 \Big( T_{\mu \nu}(x +a\hat{\nu}) 
 \nonumber \\ &&
 +  T_{\mu \nu}(x +a\hat{\mu}+a\hat{\nu})
 -  T_{\mu \nu}(x -a\hat{\nu}) -  T_{\mu \nu}(x +a\hat{\mu}-a\hat{\nu})
 \Big) 
 \nonumber\\
 &=& J_\mu(\hat{q}) + \frac{1}{2} \ccvc  
 \cos( a q_\mu/2) \sin (a q_\nu) \sum_x  e ^{iq x} T_{\mu \nu}(x) \;,
 \end{eqnarray} 
 where 
 $T_{\mu \nu}(x) \equiv \psib(x) \sigma_{\mu \nu} \psi(x)$.  
 Unfortunately it turns out that this choice introduces very large
 $\mathcal{O}(a^2 Q^2)$ errors into $\Pi$. We found that
 \begin{equation} 
 \label{goodJ} 
 J^{\rm imp}_\mu(\hat{q}) 
 \equiv J_\mu(\hat{q}) + \ccvc \sin (a q_\nu/2)
  \sum_x  e ^{iq x} T_{\mu \nu}(x)
 \end{equation} 
 was a better choice of improved current, because it makes the
  $\mathcal{O}(a^2 Q^2)$ terms much smaller, and so we will
 adopt this definition. The difference between the definitions
 (\ref{naiveJ}) and (\ref{goodJ}) is $\mathcal{O}(a^2 Q^2)$, 
 so we are free to choose whichever definition leads to the largest
 scaling region in $Q^2$ (both agree at small $Q^2$). 
 
\begin{figure}[htb]
\begin{center}
 \epsfxsize=11cm
\epsfbox{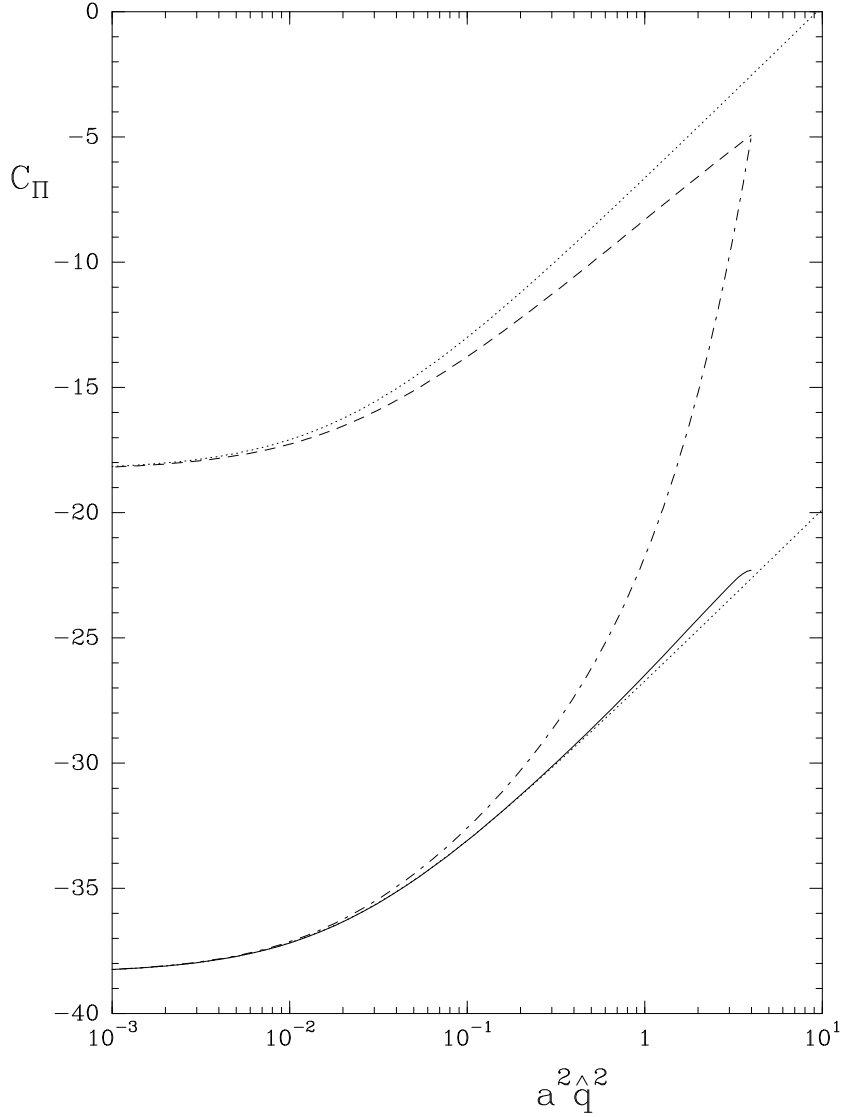}
\caption{ The effects of improving the current operator.
 We compare various curves, all calculated in the free-field
 case at the one-loop
 level with  $q \propto (0,0,0,1)$
 and $\kappa = 0.123$ (corresponding to $a m=0.065$). 
 The dotted curves show the continuum one-loop result,
 shifted vertically to match with lattice results. 
 The dashed line shows the one-loop lattice result 
 without any improvement of the current. 
 The dot-dashed curve shows the result of improving with 
 (\ref{naiveJ}) and the solid curve the result of improving
 with (\ref{goodJ}). 
    The upper two curves differ by $O(a)$, the difference 
 $~\sim a m \ln a^2 \hat q^2.$ The lower lattice curves are both
 $O(a)$ improved, but we see that the naive improvement
 (dot-dashed curve) has very large $O(a^2)$ discretisation errors. 
 Improving with  (\ref{goodJ}) produces a lattice result much closer
 to the continuum result.  
  \label{improfig}}
\end{center}
\vspace*{0.5cm}
\end{figure}

  In Fig.\ref{improfig} we show the unimproved polarisation
 tensor along with the two different choices
 of improvement term in the case of free fermions. 
 The best agreement with continuum 
 physics comes from using (\ref{goodJ}), which is the prescription
 we will use in the rest of this paper. 

In Appendix~A we give explicit expressions for 
$\Pi^{(1)}_{\mu\nu}(\hat{q})$ and  $\Pi^{(2)}_{\mu\nu}(\hat{q})$ in
terms of the link variables and the quark propagators.
 
\section{Lattice Calculation}

To facilitate the extrapolation to the chiral and continuum limits, we have
 made simulations at three different values of $\beta$ with
 three or more different
$\kappa$ values at each $\beta$. The parameters are listed in
 Table~\ref{params}. The lattice data 
for the vacuum polarisation for the individual momenta and $\beta$ and 
$\kappa$ values are given in Appendix~C.

\begin{table}[ht]
\begin{center}
\vspace{0.5cm}
\begin{tabular}{|c|c|c|c|}
 \hline
 $\beta$ &   $\kappa$  &    $V$      &   \# Conf.  \\\hline\hline
   6.0   &   0.1333    &    $16^4$   &     97      \\%\hline
   6.0   &   0.1339    &    $16^4$   &     44      \\%\hline
   6.0   &   0.1342    &    $16^4$   &     44      \\%\hline
   6.0   &   0.1345    &    $16^4$   &     44     \\\hline\hline
   6.0   &   0.1345    &    $32^4$   &     16      \\\hline\hline
   6.2   &   0.1344    &    $24^4$   &     51      \\%\hline
   6.2   &   0.1349    &    $24^4$   &     38      \\%\hline
   6.2   &   0.1352    &    $24^4$   &     51      \\\hline\hline
   6.4   &   0.1346    &    $32^4$   &     50      \\%\hline
   6.4   &   0.1350    &    $32^4$   &     29     \\%\hline
   6.4   &   0.1352    &    $32^4$   &     50      \\\hline
\end{tabular}
\vspace{0.5cm}
%\end{small}
\end{center}
\caption{Parameters of the lattice simulation. The improvement coefficient
in the fermionic action was taken to be $c_{SW}=1.769$ for $\beta=6.0$,  
$c_{SW}=1.614$ for $\beta=6.2$ and $c_{SW}=1.526$ for $\beta=6.4$~\cite{SW}.
 \label{params}}
\vspace*{0.5cm}
\end{table}

\begin{figure}[tb]
\begin{center}
\epsfig{file=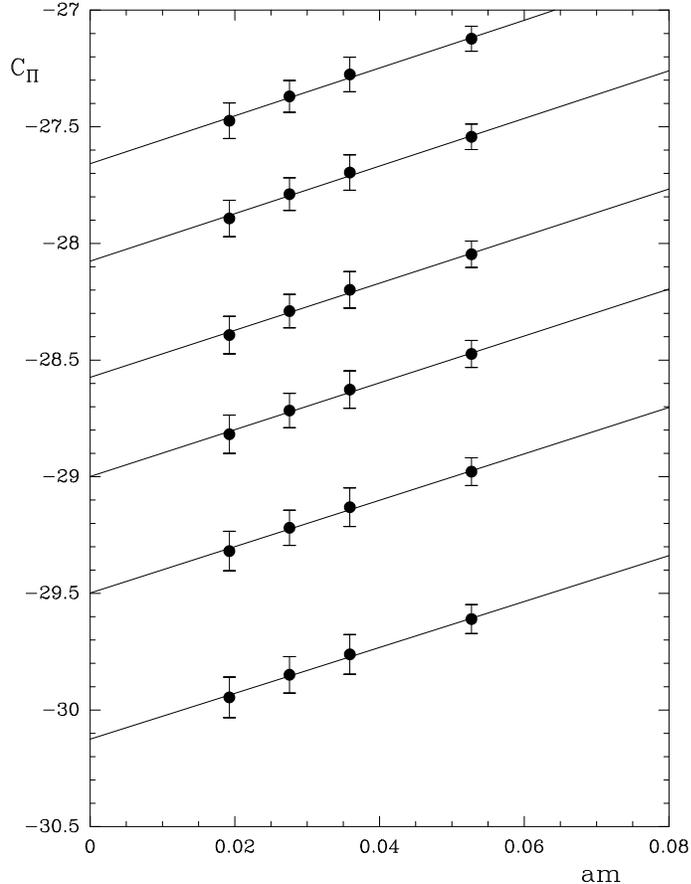,width=9cm}
 \caption{
 Chiral extrapolation of $C_\Pi(\hat{q}^2,m)$ at $\beta=6.0$. 
 The points shown range from $a^2 \hat q^2 \approx 3.0$ to $\approx 6.5$.  
 All are calculated with $\ccvc=1$. 
   \label{chiral}}
 \end{center}
 \vspace*{0.5cm}
 \end{figure}

\begin{figure}[tb] 
\begin{center}
\epsfig{file=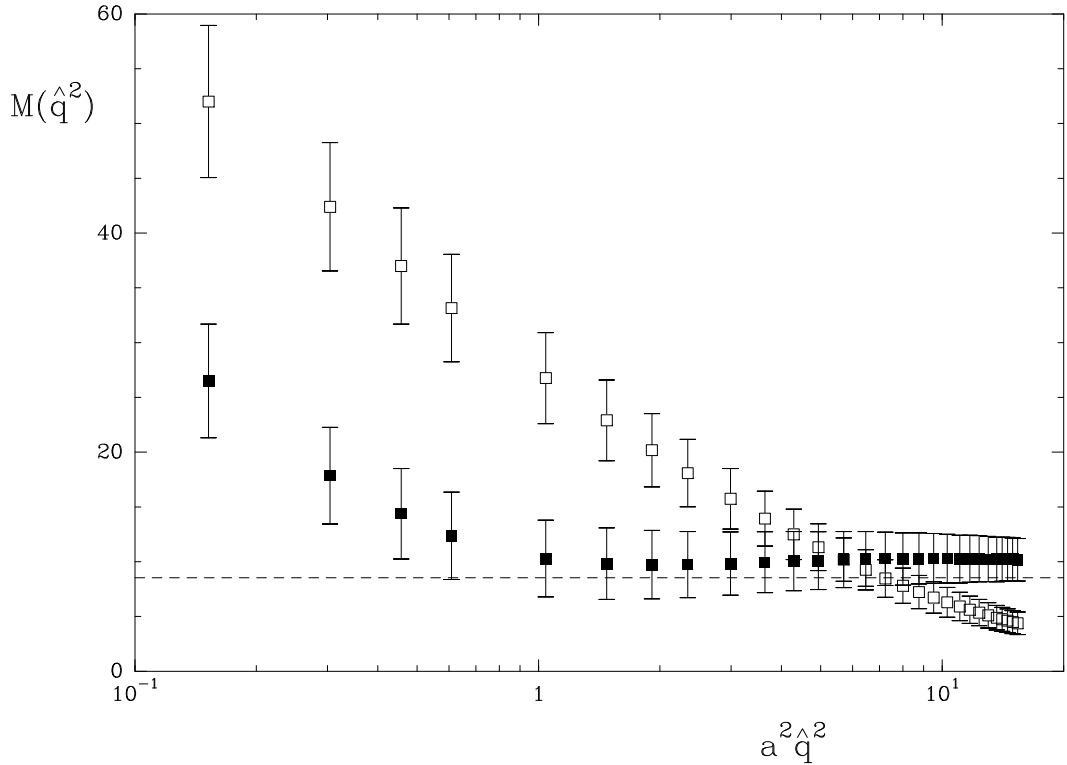,angle=270,width=14cm}
\caption{ 
The slope $M(\hat{q}^2)$ for the improved (solid points, $\ccvc=1$)
 and unimproved
(open points, $\ccvc=0$) vector current as a function of  
 $a^2\hat{q}^2$ at $\beta=6.0$.  The dashed line shows the
 height of the plateau in one-loop lattice perturbation theory,
 setting  $\ccvc=1$ in eq.(\ref{Pi_lat}). 
\label{slope} }
\end{center}
\vspace*{0.5cm}
\end{figure}

 First, let us discuss the value we use for $\ccvc$. From the fermion-loop 
contribution to the gluon propagator, computed in~\cite{Booth} to 
$\mathcal{O}(m)$ in lowest order of lattice perturbation theory, we obtain
\begin{equation}
\begin{split}
 \hspace*{-0.2cm} c_0^{(0)}(\hat{q}^2,am) = &3  \Big(\! 
\ln (a^2 \hat{q}^2) 
 -3 a m (1\! - \ccvc) \ln (a^2 \hat{q}^2)
 - 3.25275141(5) \\
 &+1.19541770(1)\, \ccvc - 7.06903716(4)\, \ccvc^2 \\ 
 &+ a m  \big( 6.46270704(30)-  5.29413266(6)\, \ccvc \\  
 &+1.67389761(2)\, \ccvc^2 \big) \Big)+\mathcal{O}(a^2, m^2) \, ,
\end{split}
\label{Pi_lat}
\end{equation}
where $am = 1/2\kappa - 1/2\kappa_c$.
 We use $c_0^{(0)}$ to refer to the lowest order, $g^0$, 
 perturbative contribution to $c_0$. 
 As said earlier, the $\hat q^2$ dependence of $\Pi$ and $c_0$ is
  physical. Therefore in an $O(a)$-improved calculation
 there should be no $O(a)$ terms which depend on  $\hat q^2$. 
 On the other hand a constant added to $c_0$ has no physical effect, 
 so there is no objection to constant terms of $O(a)$ in
 eq.(\ref{Pi_lat}). We see that there is an 
unphysical $am \ln(a^2 \hat{q}^2)$ term in (\ref{Pi_lat}) unless 
$\ccvc = 1 + \mathcal{O}(g^2)$. In the following we take $\ccvc=1$ and
make the ansatz
\begin{equation}
 C_\Pi(\hat{q}^2,am) = C_\Pi(\hat{q}^2,am=0) +am\, M(\hat{q}^2) \, .
\end{equation}
In Fig.~\ref{chiral}
 we show the quark mass dependence of $C_\Pi(\hat{q}^2,am)$ for several
 momenta, which justifies assuming a linear $am$ dependence. In 
Fig.~\ref{slope} we show the slope
$M(\hat{q}^2)$ with and without improvement of the conserved vector current.
 The derivative $M$ should tend to a constant when $Q^2 \gg m^2$. 
 With no improvement term (open points) we see that $M$ has a
 logarithmic dependence on $\hat{q}^2$, corresponding to an 
  unphysical $a m \ln(a^2\hat{q}^2)$ term in $c_0$. 
 We see that for $\ccvc=1$ the slope is, within error bars, independent 
 of $\hat{q}^2$ down to small momenta. This shows that the 
 choice $\ccvc=1$ has eliminated or greatly reduced the 
 unphysical logarithm in $c_0$. 
 We conclude that the tree-level value $\ccvc=1$ is a good
 choice
\footnote{If $\ccvc$ should
ever be computed non-perturbatively, or in perturbation theory, it should be
kept in mind that we have used eq.~(\ref{goodJ}) as the definition
 of the improvement term.}.

\begin{figure}[tb] 
\begin{center}
\epsfysize=11cm
\epsfbox{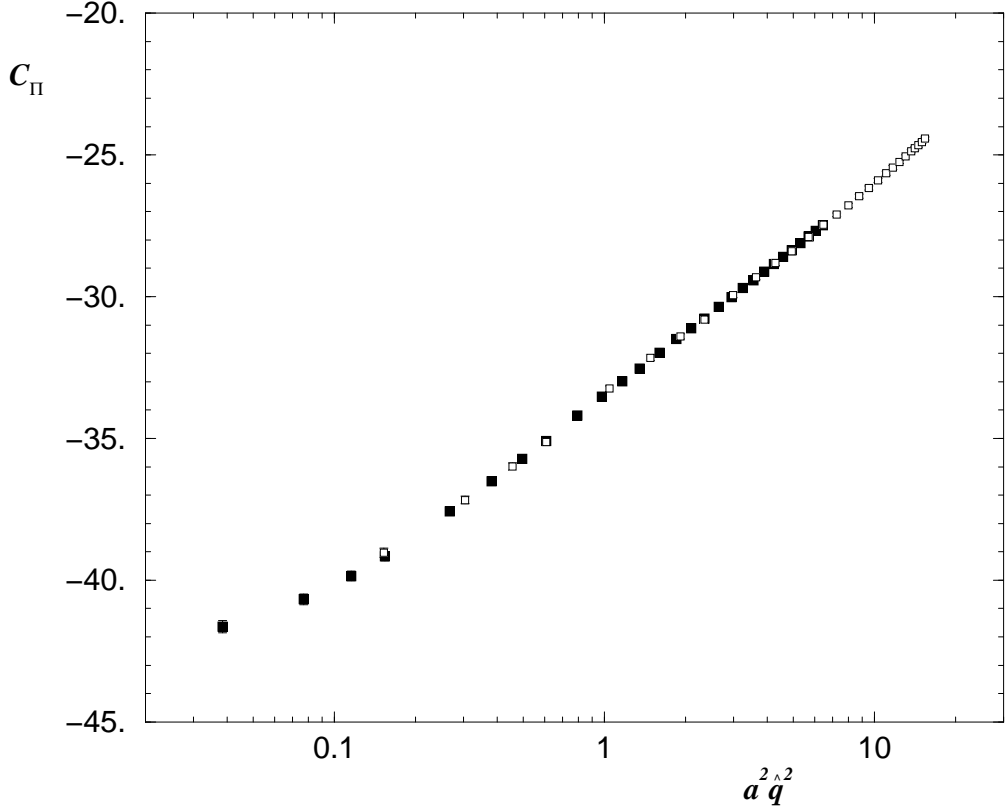}
\caption{ A check for finite volume effects. 
At  $\beta=6.0$, $\kappa=0.1345$  we have measured the polarisation
 tensor both on a $16^4$ lattice (white points) and on a $32^4$ 
 lattice (black points). The agreement is excellent, and we conclude
 that finite volume effects are negligible. 
  \label{finsize}}
\end{center}
\vspace*{0.5cm}
\end{figure}

 The lattice sizes in Table~\ref{params}
 were chosen such that the physical volume is approximately
 equal for all three $\beta$ values.
 As  we have done simulations at
 $\beta=6.0$, $\kappa=0.1345$ on two different lattice volumes
 we check for finite volume 
 effects. We have not found any, see Fig.\ref{finsize}. This figure
 also shows that on the larger lattice we can measure the vacuum 
 polarisation at much lower values of $Q^2$, which is an important 
 advantage.  The quark boundary conditions are antiperiodic in all
 four directions, while gluon and photon fields have periodic
 boundary conditions in all directions.

 \section{Comparison with perturbation theory} 

\begin{figure}[tb] 
\begin{center}
\epsfig{file=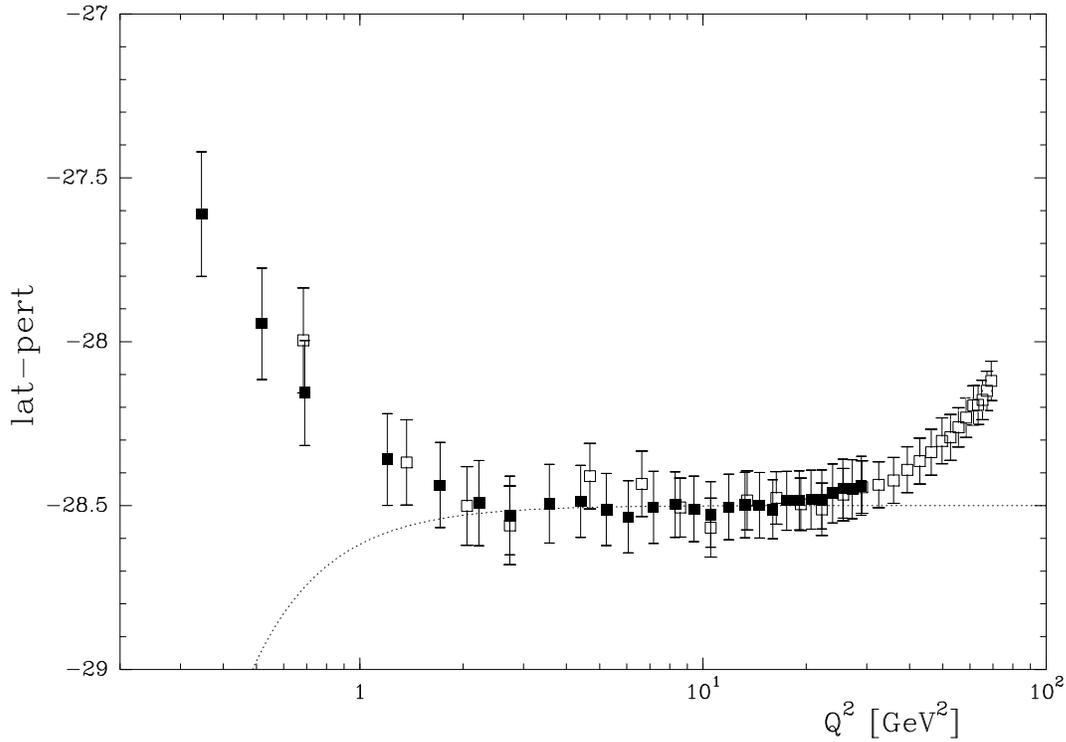,angle=270,width=14cm}
\caption{ The deviation of lattice data from continuum perturbation theory. 
 The data are at  $\beta=6.0$, $\kappa=0.1345$ 
  on a $16^4$ lattice (white points) and on a $32^4$ 
 lattice (black points). Below $Q^2 \sim 2 \gev^2$ 
 there is a visible deviation from perturbation theory. 
 The dotted line shows the effect expected from a gluon
 condensate as expected in the OPE. 
  \label{deviate}}
\end{center}
%\vspace*{0.5cm}
\end{figure}

   The first thing to do is to compare lattice results with continuum
 perturbation theory, which we do in Fig.\ref{deviate}.
 For the perturbative contribution
 $c_0^{\rm pert}(\hat{q}^2, m)$ we use the renormalisation-group
 improved result given in eqs.~(\ref{improved}) and (\ref{c0massive})
 of Appendix~B. We use  $\Lambda_{\overline{MS}} = 243 \, 
\mbox{MeV}$~\cite{Booth,Alpha} and $\mu = 1/a$, and we identify $Q^2$ with 
$\hat{q}^2$. The $r_0$ parameter is used to fix the scale~\cite{Sommer}, 
with $r_0 = 0.5 \, \mbox{fm}$. 

 $C_\Pi$ calculated on the lattice and $C_\Pi$
 in the continuum can differ by an 
 integration constant which can depend on $\mu$ and $a$. In lowest order
 perturbation theory this constant is found by comparing $c_0^{(0)}$ in
 $\overline{MS}$, eq.~(\ref{ggz1}), with  $c_0^{(0)}$ calculated in
 lattice perturbation theory, eq.~(\ref{Pi_lat}). Setting $\ccvc$ equal
 to 1 and $\mu = 1/a$ we find that 
 $\Delta c_0 = c_0^{\rm lat}-c_0^{\overline{MS}} = -22.379$.
 In next to leading order this becomes 
 \begin{equation}
 \Delta c_0 =  - 22.379
 -12/11 \,\ln(\alpha^{\overline{MS}}/\alpha^{\rm lat})
 + \mathcal{O}(\alpha^{\rm lat}).
 \end{equation}
 In view of this result, the value $\Delta c_0 \approx -28.5$ 
 seen in Fig.\ref{deviate} is reasonable. 

  Fig.\ref{deviate} shows that the lattice results deviate from
 perturbation theory at large $Q^2$. This deviation,
 which sets in at $a^2 \hat{q}^2 \sim 5$, is probably a sign of
 $O(a^2)$ lattice artefacts.

 More interesting are the deviations from perturbation theory
  at low $Q^2$. These are especially large
 at low quark mass. The OPE, eq.(\ref{OPE}),
 would suggest that the effects of a gluon condensate should 
 show up at small $Q^2$. In Fig.\ref{deviate} we show
 the OPE prediction for a gluon condensate 
  $\frac{\alpha}{\pi}\langle G G \rangle  = 0.012\gev^4$. 
  The deviations which we see in the data 
 do not look like the effects expected from a gluon condensate. 
 The sign is the opposite of what the OPE predicts, and the deviation
 is probably not growing as quickly as $1/Q^4$. 

 \section{A simple model of the vacuum polarisation at low $Q^2$}

    We have seen that perturbation theory, even when supplemented
 with higher twist terms from the operator product expansion, has
 difficulty in explaining the low $Q^2$ region of the data. 
 Can we understand this region in some other way? 

 The cross section ratio $R(s)$ is given by the cut in
 the vacuum polarisation, or in other words there are dispersion
 relations which give the vacuum polarisation if we know $R(s)$. 

\begin{table}[ht]
\begin{center}
\vspace{0.5cm}
\begin{tabular}{|c|c|c|c|c|}
 \hline
$\beta$& $\kappa$&$a m_{PS}$  & $a m_V$  & $1/f_V$ \\\hline\hline
  6.0  & 0.1333  & 0.4122(9)  & 0.5503(20) & 0.2055(14) \\
  6.0  & 0.1339  &{\it 0.3381(15)}&{\it 0.5017(40)}& {\it 0.2217(15)}  \\
  6.0  & 0.1342  & 0.3017(13) & 0.4904(40) & 0.2293(14)  \\
  6.0  & 0.1345  &{\it 0.2561(15)}&{\it 0.4701(90)}& {\it 0.2387(90)}
  \\\hline\hline
  6.2  & 0.1344  & 0.3034(6) & 0.4015(17) & 0.2210(20)  \\
  6.2  & 0.1349  & 0.2431(6) & 0.3663(27) & 0.2403(24)  \\
  6.2  & 0.1352  & 0.2005(9) & 0.3431(60) & 0.2474(47)  \\
 \hline\hline
  6.4  & 0.1346  & 0.2402(8) & 0.3107(16) & 0.2252(21)  \\
  6.4  & 0.1350  & 0.1933(7) & 0.2800(20) & 0.2423(19) \\
  6.4  & 0.1352  &{\it 0.1661(10)}&{\it 0.2613(40)}& {\it 0.2448(35)} \\
 \hline
\end{tabular}
\vspace{0.5cm}
\end{center}
\caption{Lattice data on the vector meson used as input
 for the dispersion relation fits~\cite{scaling,quenchspectro}.
 Numbers in italics have been interpolated from nearby $\kappa$ values.
 \label{rhodata} }
\vspace*{0.5cm}
\end{table}

    One way of modelling $\Pi$ is to make a model for $R(s)$, 
 and then calculate $\Pi$ from $R$, using the dispersion
 relation eq.(\ref{dispersion}). This should be quite robust, 
 since $\Pi$ gets contributions from a large range in $s$, little
 inaccuracies in the model $R$ get washed out, and we can hope that
 even a crude representation of $R$ will give a good result for $\Pi$. 
 We will keep the model simple so that we can do all the integrals
 analytically. 

  Perturbatively
 \begin{equation} 
 R(s) = \sum_f e_f^2 N_c \left(1 + \frac{\alpha_s}{\pi} + \cdots \right)  
 \label{Rpert1} 
 \end{equation} 
 where $N_c$ is the number of colours (3 in our case). 
 We know that really the low $s$ behaviour of $R$ is more complicated than 
 that, it is dominated by the $\rho(770)$, $\omega(782)$ 
 and $\phi(1020)$ mesons. 
 Following~\cite{SVZII}, let us make the following model for 
 $R$. We ignore the splitting between the $\rho$ and $\omega$, 
  which comes from the $A_\Pi$-type diagrams which we have dropped. 
 We  also treat these mesons as narrow resonances, each 
 contributing a $\delta$ function to $R$. The continuum part of $R$
 takes a while to 
 climb up to the value in eq.(\ref{Rpert1}). We will represent this 
 rise by a step  function at some value $s_0$. So, our model is 
 \begin{equation} 
 R(s) =\sum_f e_f^2 \left( A \delta(s - m_V^2)
  + B \Theta(s -s_0) \right)   
 \label{model} 
 \end{equation} 
 where we would expect $B$ to be slightly above 3 in order to
 match eq.(\ref{Rpert1}). 

 Using the dispersion relation, this $R(s)$ translates into a vacuum
 polarisation
 \begin{equation} 
 C_\Pi(Q^2) = B \ln (a^2Q^2 + a^2s_0) - A/(Q^2 + m_V^2) + K
 \label{disp_Ansatz} 
 \end{equation} 
 where $K$ is a constant which is not determined from the dispersion
 relation, and which never appears in any physical quantity. 
 Again we identify the continuum quantity $Q^2$ with the lattice
 quantity $\hat q^2$. 

  The constant $A$ can be expressed in terms of the decay 
 constant $f_V$, which has been measured on the lattice. 
 The cross section for the production of a
 narrow vector resonance, $V$, 
 is~\cite{SVZII}
 \begin{equation} 
 \sigma_{e^+e^-\rightarrow V}(s) = 12 \pi^2 \delta(s - m_V^2)
  \frac{\Gamma_{V\rightarrow e^+e^-}}{m_V} \;.
 \end{equation} 
 The partial width $\Gamma_{V\rightarrow e^+e^-}$ is 
 related to a meson decay constant $g_V$~\cite{DecayDef,SVZII} by 
 \begin{equation}
  \Gamma_{V\rightarrow e^+e^-} = 
 \left(\frac{4 \pi \alpha_{em}^2}{3}\right) \frac{m_{V}}{g_V^2}\;, \\
 \end{equation}
  where
 \begin{equation}
 \langle 0 | J^{em}_\mu | V, \varepsilon \rangle
 = \varepsilon_\mu \frac{m_V^2}{g_V}. 
 \end{equation}
 Here $\varepsilon_\mu$ is the polarisation vector of the meson. 
 On the lattice it is more natural to define decay constants
 $f_V$ in terms of currents with definite isospin~\cite{scaling}
 \begin{equation}
  \langle 0 | \mathcal{V}_\mu | V, \varepsilon \rangle
 = \varepsilon_\mu \frac{m_V^2}{f_V} 
 \end{equation}
  with 
 \begin{equation}
 \mathcal{V}_\mu^{I=1} = \frac{1}{\sqrt{2} } 
 \left( \bar{u} \gamma_\mu u - \bar{d} \gamma_\mu d \right) 
 \end{equation} 
  for the $\rho^0$ and 
 \begin{equation}
 \mathcal{V}_\mu^{I=0} = \frac{1}{\sqrt{2} } 
 \left( \bar{u} \gamma_\mu u + \bar{d} \gamma_\mu d \right) 
 \end{equation} 
  for the $\omega$
 (ignoring any ${\bar s}s$ admixture in the $\omega$). 
 The relationship between the two definitions is 
 \begin{eqnarray}
 \frac{1}{g_\rho} &=& \frac{e_u -e_d}{\sqrt{2}} \;  \frac{1}{f_\rho} 
          = \frac{1}{\sqrt{2}} \; \frac{1}{f_\rho} \;, \\
 {}\nonumber \\ 
 \frac{1}{g_\omega} &=& \frac{e_u + e_d} {\sqrt{2}}\; \frac{1}{f_\omega}
  = \frac{1}{3\sqrt{2}}\; \frac{1}{f_\omega} \;. 
 \end{eqnarray} 

  In terms of these decay constants 
 \begin{eqnarray} 
 R(s) &=& \sum_V  12 \pi^2 \delta(s -m_V^2) \frac{m_V^2}{g_V^2}
  + {\rm continuum} \\
 &=& 12 \pi^2 \delta(s-m_\rho^2) \frac{m_\rho^2}{f_\rho^2}
 \frac{(e_u-e_d)^2}{2}  \nonumber \\
 && + 12 \pi^2 \delta(s-m_\omega^2) \frac{m_\omega^2}{f_\omega^2}
 \frac{(e_u+e_d)^2}{2} + {\rm continuum.} 
 \label{R_resonance}
 \end{eqnarray} 
 Neglecting annihilation diagrams implies that $m_\omega = m_\rho$
 and $f_\omega = f_\rho$. Experimentally both relations are  
 fairly accurate. The mass ratio $m_\omega / m_\rho$ is 1.02.
 $f_\omega = f_\rho$ implies 
  $\Gamma_{\omega \rightarrow e^+e^-} = \frac{1}{9} 
 \Gamma_{\rho^0 \rightarrow e^+e^-}$, while the experimental ratio is 
 $0.089(5)$~\cite{PDG}. Equating the mass and decay constant of
 the $\omega$ and $\rho$ in eq.~(\ref{R_resonance}) gives
 \begin{equation} 
 R(s) = 12 \pi^2 \delta(s-m_\rho^2) \frac{m_\rho^2}{f_\rho^2}
 \left(e_u^2 + e_d^2 \right) + {\rm continuum} \;;
 \end{equation} 
 so 
 \begin{equation}
 A = 12 \pi^2 \frac{m_V^2}{f_V^2}. 
 \label{Aval} 
 \end{equation} 

 \section{Dispersion relation fits} 

  We first try making fits to the lattice data for the
 vacuum polarisation $C_\Pi$ using eq.~(\ref{disp_Ansatz})
 with $B, K$ and $s_0$ as free parameters.
 To avoid problems from lattice artefacts of
 $\mathcal{O}(a^2  \hat{q}^2)$ we have only used data with
  $a^2 \hat{q}^2 < 5$.  We will call this simple ansatz Fit~I.  
 $A$, the weight of the vector meson contribution, is determined
 by eq.~(\ref{Aval}).  The vector meson masses and decay constants
 which we use are shown in Table~\ref{rhodata}. They have
 been taken from~\cite{scaling,quenchspectro}.  
 We can compare the values for $B$ and $K$ with the one-loop
 lattice perturbation theory result, eq.~(\ref{Pi_lat}), which gives 
 $B=3$ and $K = -27.38 + a m \; 8.53$. 
\begin{table}[ht]
\begin{center}
\vspace{0.5cm}
\begin{tabular}{|c|c|c|c|c|c|c|}
 \hline
 $\beta$ & $\kappa$ & $V$ & $a^2 s_0$ & $B$    & $K$   & $\chi^2$ \\\hline\hline
 6.0  & 0.1333  & $16^4$ & 0.395(32) & 3.13(6) & -32.98(11)& 4.3 \\
 6.0  & 0.1339  & $16^4$ & 0.397(43) & 3.13(8) & -33.14(14)& 2.2 \\
 6.0  & 0.1342  & $16^4$ & 0.391(39) & 3.10(8) & -33.18(13)& 3.0\\
 6.0  & 0.1345  & $16^4$ & 0.409(61) & 3.11(9) & -33.30(16)& 2.5 \\\hline
 6.0  & 0.1345  & $32^4$ & 0.403(79) & 3.12(10) & -33.30(18)& 0.7 \\\hline
\hline
 6.2  & 0.1344  & $24^4$ & 0.282(21) & 3.10(5) & -32.94(7)& 1.9\\
 6.2  & 0.1349  & $24^4$ & 0.278(24) & 3.07(5) & -33.06(8)& 1.9 \\
 6.2  & 0.1352  & $24^4$ & 0.253(25) & 3.06(5) & -33.10(7)& 2.2\\
 \hline\hline
 6.4  & 0.1346  & $32^4$ & 0.164(13) & 3.06(3) & -32.68(5)& 0.8\\
 6.4  & 0.1350  & $32^4$ & 0.150(14) & 3.03(4) & -32.76(6)& 0.6 \\
 6.4  & 0.1352  & $32^4$ & 0.124(12) & 3.02(3) & -32.78(5)& 1.2 \\\hline
\end{tabular}
\vspace{0.5cm}
\end{center}
\caption{ The result of Fit~I, eq.~(\ref{disp_Ansatz}),
 including only data with $a^2 \hat{q}^2 < 5$.
 \label{fit1}}
\vspace*{0.5cm}
\end{table}

\begin{figure}[htb] 
\begin{center}
\epsfig{file=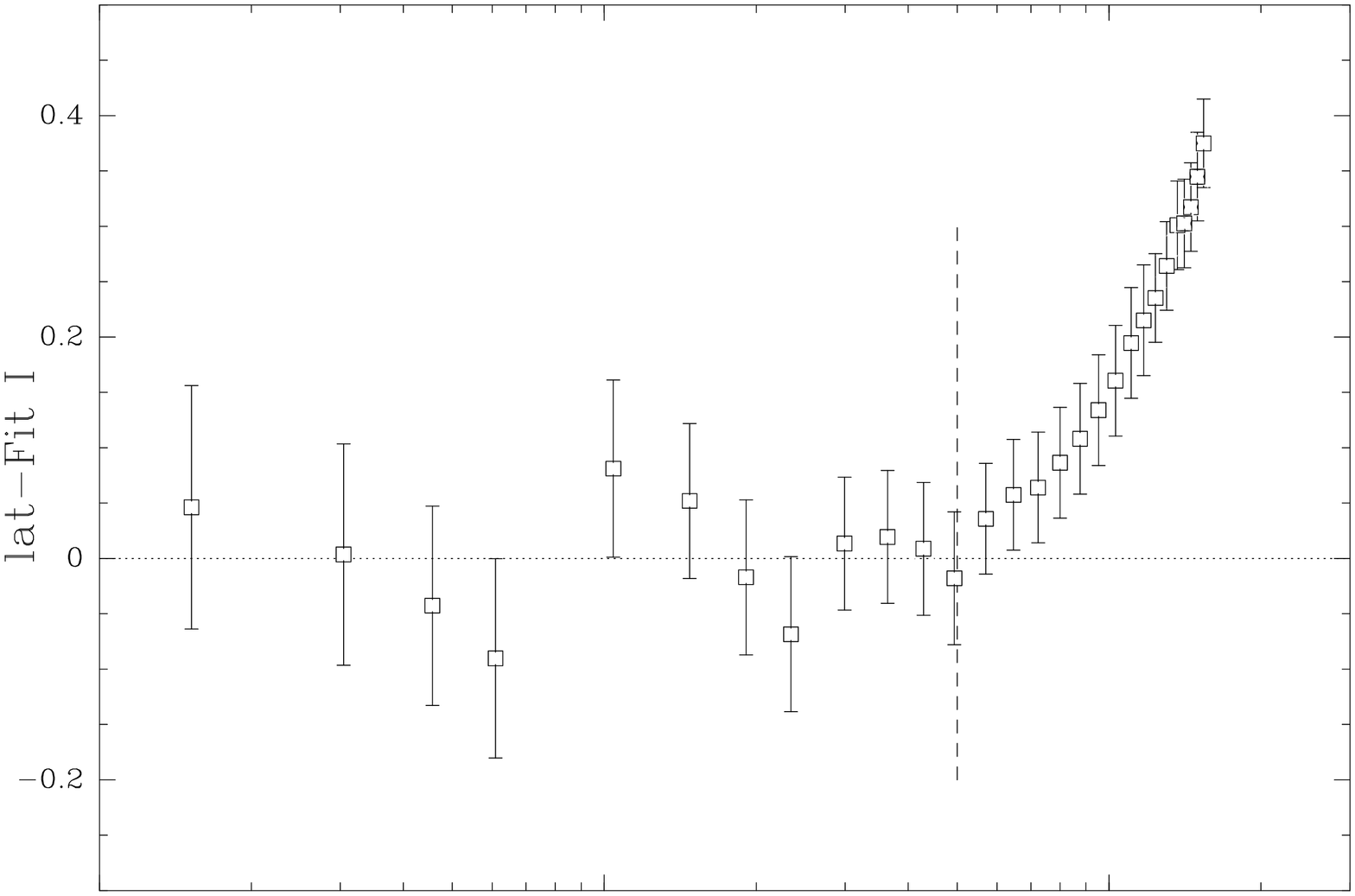,angle=270,width=14cm}
\vspace*{-5.5mm}{
\epsfig{file=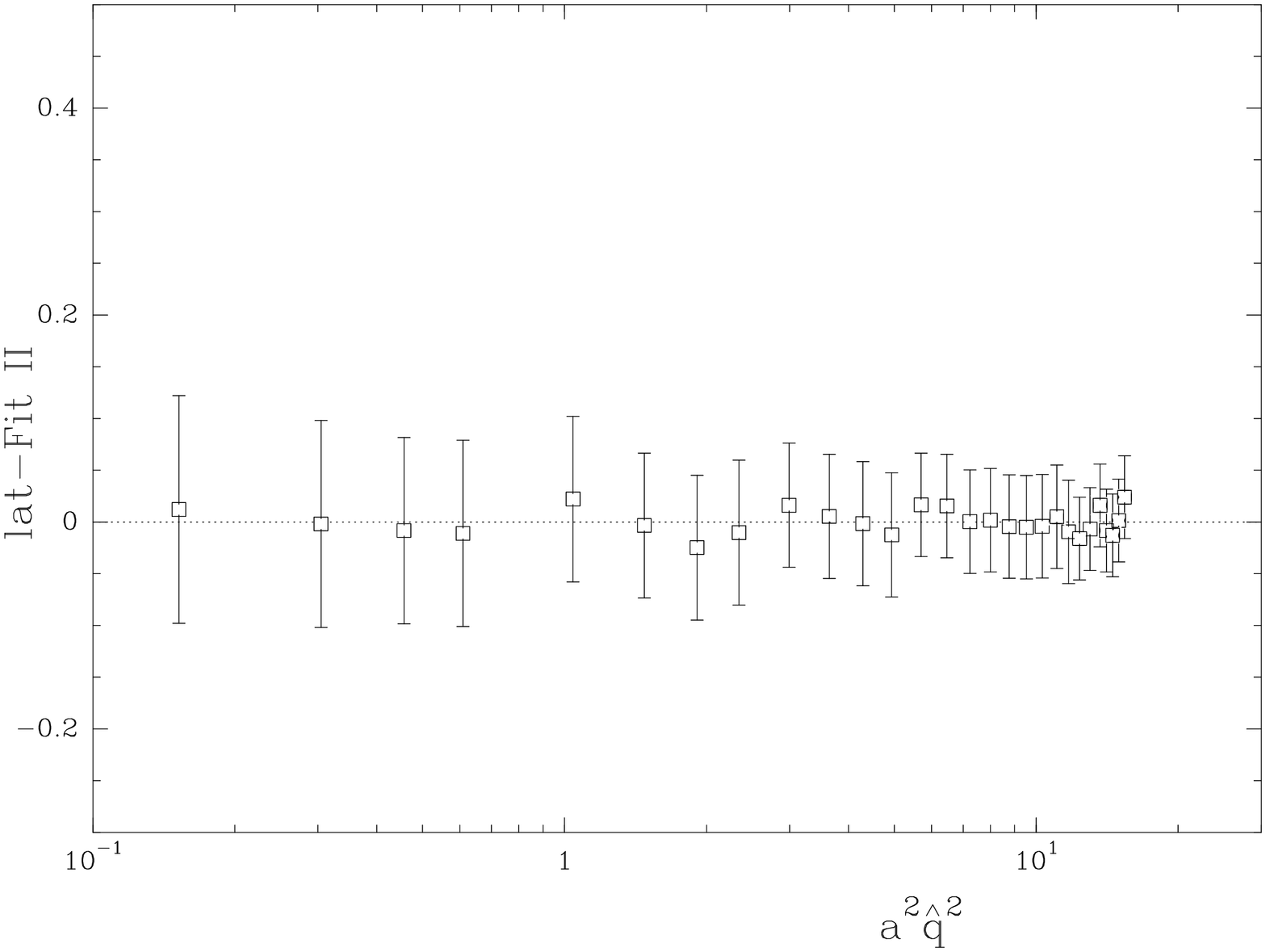,angle=270,width=14cm} }
\caption{ A comparison between Fit~I and Fit~II.
 The data are at  $\beta=6.0$, $\kappa=0.1333$ on a $16^4$ lattice.
 Fit~I only uses data with 
 $a^2 \hat{q}^2 <5$ (vertical dashed line). The second fit describes the data
 better.
  \label{fit12}}
\end{center}
 \vspace*{0.5cm}
\end{figure}

 A typical result is shown in the top panel of Fig.\ref{fit12}. 
 As can be seen from Table~\ref{fit1}, the $\chi^2$ values for the fits 
 are low in every case. Nevertheless we see that the data deviates
 quite strongly from the fit when $a^2 \hat{q}^2$ is large. The 
 small deviations from the fit at smaller $ \hat{q}^2$ are not
 random - they occur in the same place in every data set, 
 and depend on the direction of $q$. Points where $q$ is near
 the diagonal direction $(1,1,1,1)$ lie lower than points 
 measured for momenta away from this diagonal direction. 

 To correct for this behaviour we add terms to our fit to parameterise
 $\mathcal{O}(a^2)$ lattice errors. There are two possible terms at
  $\mathcal{O}(a^2)$: 
 \begin{equation} 
 a^2 q^2 {\rm \quad and \quad} a^2 \frac{\sum_\mu q_\mu^4}{q^2} \;. 
 \end{equation}
 The second term has only cubic symmetry, and can give rise to 
 a dependence of $C_\Pi$ on the direction of $q$, which would not
 occur in the continuum. To include these terms we fit with the ansatz
 (Fit~II)

 \clearpage

 \clearpage \noindent
 \begin{equation} 
 C_\Pi(\hat{q}^2) = B \ln(a^2\hat{q}^2 + a^2 s_0)
 -\frac{A}{\hat{q}^2 +m_V^2} + K + 
 U_1 a^2 \hat{q}^2 + U_2 h(q)
 \label{Fit2} 
 \end{equation} 
 where
  \begin{equation}
  h(q) = \frac{ \sum_\mu \sin^4 a q_\mu
    - \frac{1}{4} \left( \sum_\mu \sin^2 a q_\mu \right)^2}{ a^2 \hat{q}^2}
  \approx  a^2 \left[ \frac{\sum_\mu q_\mu^4}{q^2}
  - \frac{1}{4}q^2 \right] \;.
  \end{equation}

\begin{figure}[htb] 
  \vspace*{-0.1cm}
\begin{center}
\epsfig{file=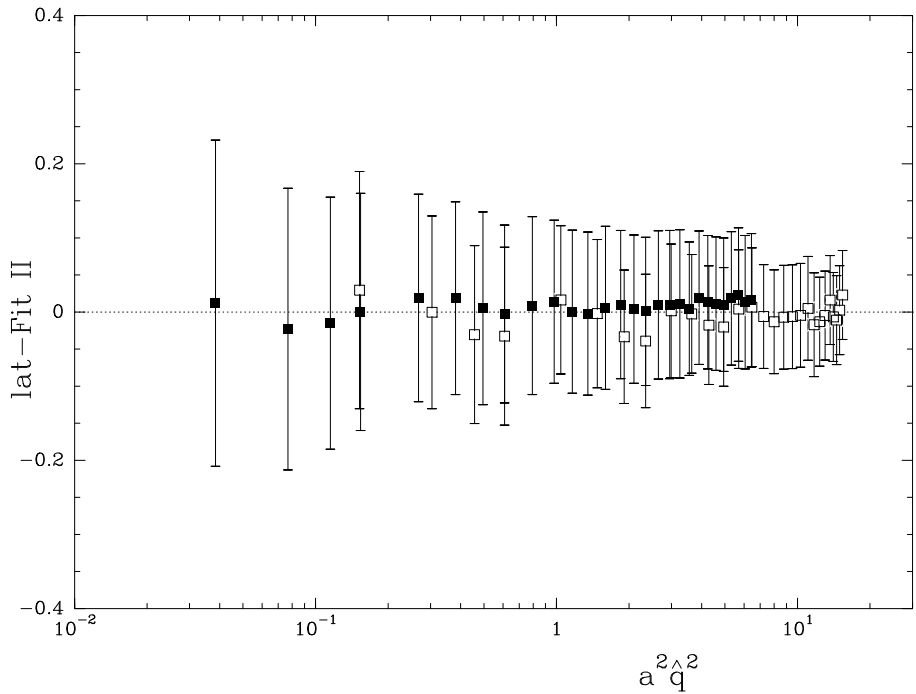,angle=270,width=12cm}
\caption{ The deviation of lattice data from 
 the fit of Table~\ref{fit2}. 
 The data are at  $\beta=6.0$, $\kappa=0.1345$ 
  on a $16^4$ lattice (white points) and on a $32^4$ 
 lattice (black points). The fit describes the data
 extremely well. 
  \label{gemfit}}
\end{center}
\end{figure}

\begin{table}[ht]
\begin{center}
 \vspace*{1.5cm}
\begin{tabular}{|c|c|c|c|c|c|c|}
 \hline
 $\beta$ & $\kappa$ &$V$& $a^2 s_0$ & $B$    & $K$   & $\chi^2$ \\\hline\hline
 6.0  & 0.1333 &$16^4$ & 0.357(50) & 3.00(12) & -32.97(17)& 1.52 \\
 6.0  & 0.1339 &$16^4$ & 0.357(68) & 2.99(16) & -33.13(23)& 0.69 \\
 6.0  & 0.1342 &$16^4$ & 0.354(62) & 2.97(15) & -33.18(21)& 0.83 \\
 6.0  & 0.1345 &$16^4$ & 0.37(10) & 2.97(17) & -33.30(25)& 0.70 \\\hline
 6.0  & 0.1345 &$32^4$ & 0.35(11) & 2.92(33) & -33.22(28)& 0.06 \\\hline
 6.0  & 0.1345 &$16^4\&32^4$ & 0.358(77) & 2.93(11) & -33.22(15)& 1.49
 \\\hline\hline
 6.2  & 0.1344 &$24^4$ & 0.262(35) & 3.00(12) & -32.94(12)& 0.14 \\
 6.2  & 0.1349 &$24^4$ & 0.252(40) & 2.94(13) & -33.05(14)& 0.12 \\
 6.2  & 0.1352 &$24^4$ & 0.232(38) & 2.95(12) & -33.11(12)& 0.13 \\\hline\hline
 6.4  & 0.1346 &$32^4$ & 0.156(21) & 3.01(11) & -32.69(7)& 0.08\\
 6.4  & 0.1350 &$32^4$ & 0.144(23) & 2.99(12) & -32.78(9)& 0.10 \\
 6.4  & 0.1352 &$32^4$ & 0.117(18) & 2.97(10) & -32.81(7)& 0.11 \\\hline
\end{tabular}
\vspace*{0.5cm}
\end{center}
\caption{ The result of Fit~II, including all data. 
 \label{fit2}}
\end{table}

  The function $ h(q)$ has been chosen so that it vanishes when 
  $q \propto (1,1,1,1)$, as most of our momenta are near this direction. 
  Looking at the lower panel of  Fig.~\ref{fit12} we see that the 
  $U_1$ term does a good job of describing the high $\hat{q}^2$ data, 
  while the $U_2$ term successfully describes the direction dependence
  of the data. Fits with the ansatz~(\ref{Fit2}) show
 practically no deviation from the data. This can also be seen in
 Fig.\ref{gemfit}, where we show the deviation in the case
 where we have data on two lattice sizes. 

\begin{figure}[htb] 
 \vspace*{0.5cm}
\begin{center}
\epsfig{file=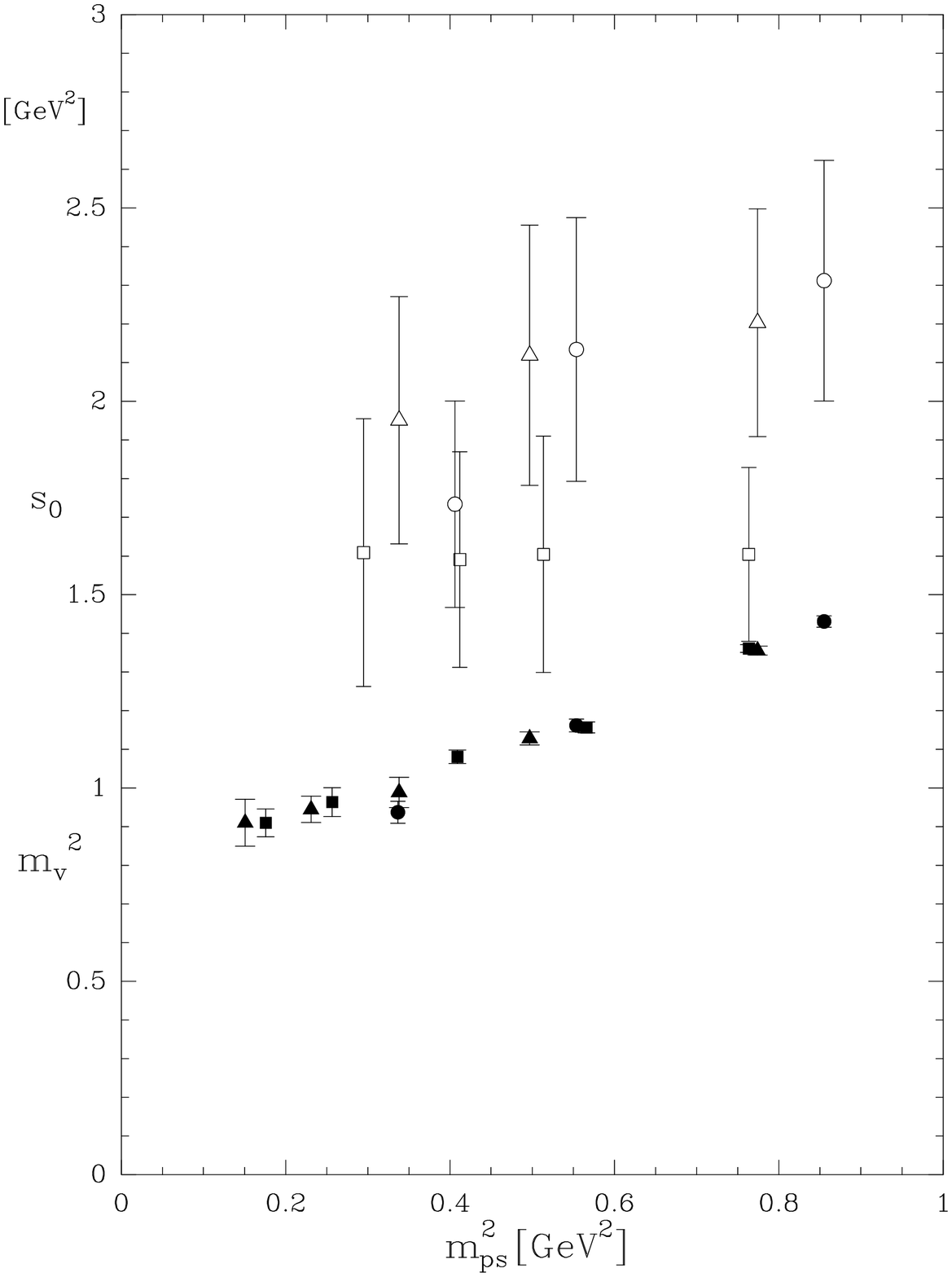,width=11cm}
\caption{ The threshold $s_0$ (white points)
 compared with $m_\rho^2$ (black points). 
 Squares are for $\beta=6.0$, triangles for 6.2, and 
 circles for 6.4. 
  \label{s0pic}}
\end{center}
 \vspace*{0.5cm}
\end{figure}

\begin{figure}[htb] 
\begin{center}
\epsfig{file=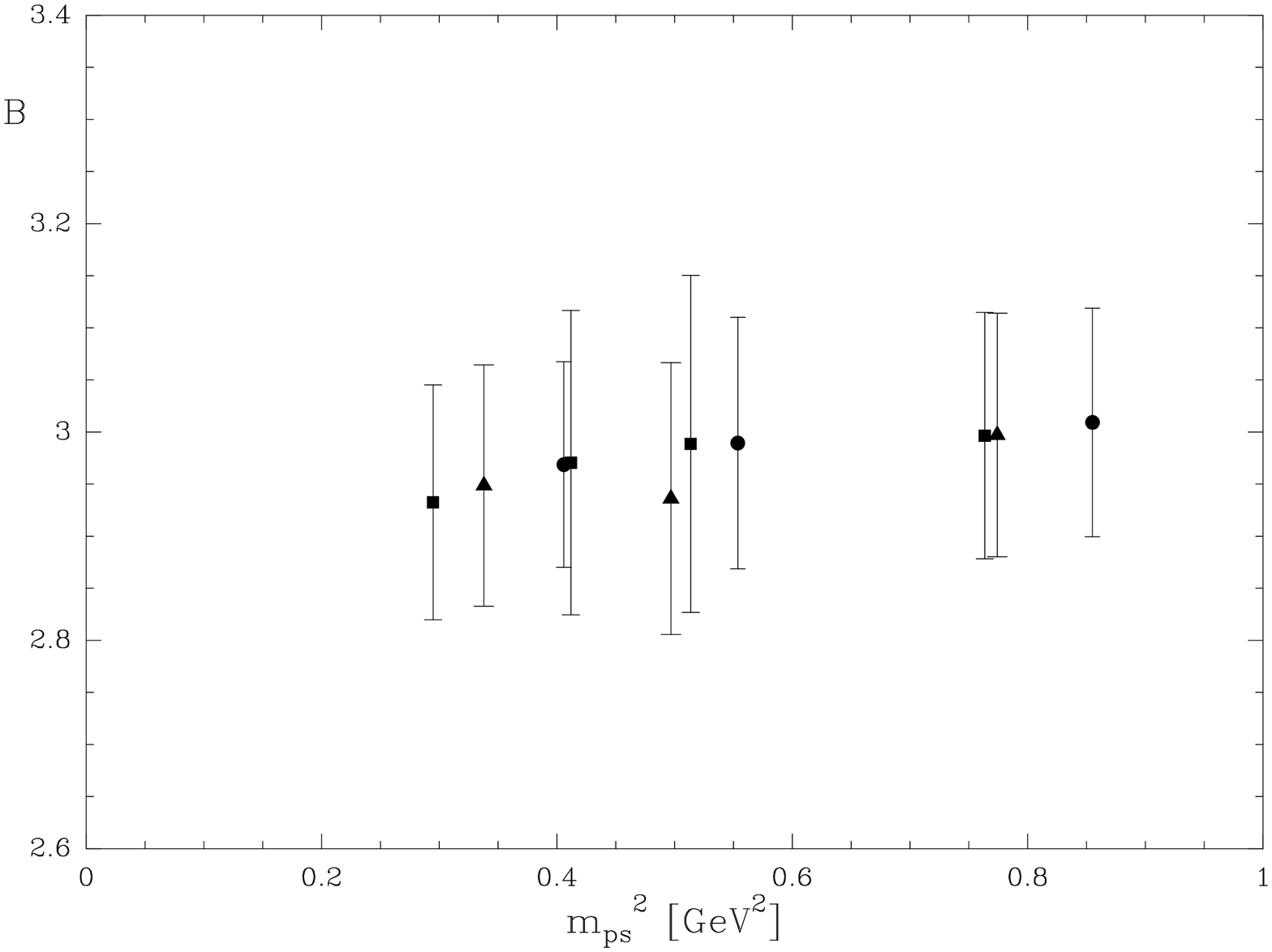,angle=270,width=13cm}
\caption{ The parameter $B$ from Table~\ref{fit2}. 
 The $\beta$ values are shown by the same symbols as in Fig.\ref{s0pic}. 
  \label{bpic}}
\end{center}
\end{figure}

 In Fig.\ref{s0pic} we show the results converted into 
 physical units. The agreement between the different $\beta$
 values is fair, and the value of $s_0$ in agreement with
 phenomenology (for example,~\cite{Bertlmann} finds 
 $s_0 = 1.66(22)\gev^2$ in the I=1 channel).  

 In Fig.\ref{bpic} we show the values for the fit parameter $B$. 
 The value we find is always very close to the tree-level value 3.

 \section{The anomalous magnetic moment of the muon}

   With a good description of the vacuum polarisation tensor 
 in the low $Q^2$ region we can make statements about
  phenomenologically interesting quantities such as the contribution
 of the $u,d$ and $s$-quarks to the muon anomalous magnetic
 moment~\cite{Blum}. 

    Traditionally the hadronic contribution to the muon's anomalous
 magnetic moment is found from $R(s)$ via a dispersion
 relation (see~\cite{Nyf} for a review)
 \begin{equation}
 a_\mu^{had} = 
 \frac{\alpha_{em}^2}{3 \pi^2} \int_{4 m_\pi^2}^\infty
 \frac{ds}{s} K\left(\frac{s}{m_\mu^2}\right) R(s) 
 \label{Rint}
 \end{equation} 
 where
 \begin{equation}
  K\left(\frac{s}{m_\mu^2}\right)
  = \int_0^1 \frac{ x^2(1-x)}{x^2 +(1-x) s/m_\mu^2} \;. 
 \end{equation} 
 The muon mass, $m_\mu$, is $105.7 \mev$. 
  For us it is more useful to deform the contour and find this same
 number from the vacuum polarisation at spacelike momenta~\cite{Blum}
 \begin{eqnarray}
  a_\mu^{had} &= &
 \frac{\alpha_{em}^2}{3 \pi^2}
 \int_0^\infty \frac{d Q^2}{Q^2} F\left(\frac{Q^2}{ m_\mu^2}\right)
 12 \pi^2 \left[ \Pi(0) - \Pi(Q^2) \right]  \nonumber \\ 
   &= &
 \frac{\alpha_{em}^2}{3 \pi^2} \sum_f e_f^2 
 \int_0^\infty \frac{d Q^2}{Q^2} F\left(\frac{Q^2}{ m_\mu^2}\right)
 \left[ C_\Pi(Q^2,m_f) - C_\Pi(0,m_f) \right] \\
 &&\hspace{6cm}\mbox{} + \; {\rm annihilation} \;, \nonumber
 \end{eqnarray}
 where the kernel $F$ is 
 \begin{equation} 
 F\left(\frac{Q^2}{ m_\mu^2}\right) = \frac{16 \; m_\mu^4}{ (Q^2)^2 \; 
  \left( 1 +\sqrt{1 +4 m_\mu^2/Q^2}\  \right)^4
  \; \sqrt{1 +4 m_\mu^2/Q^2 } }\;. 
 \end{equation} 

\begin{figure}[htb] 
\begin{center}
\epsfig{file=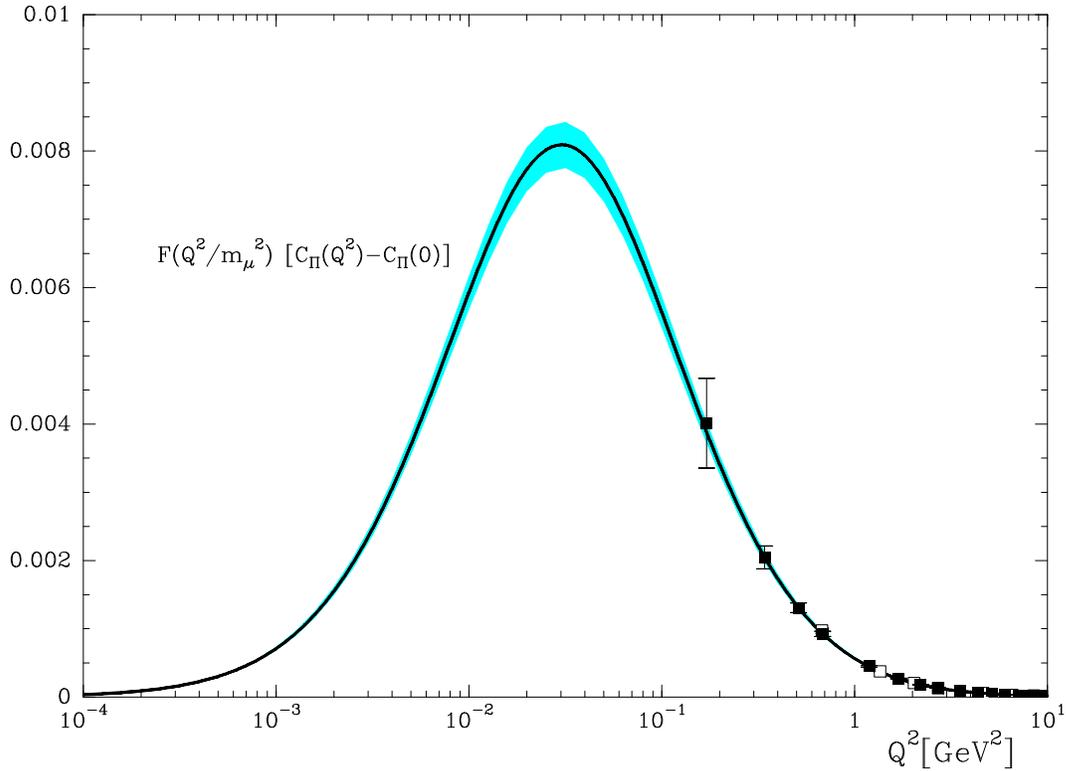,angle=270,width=14cm}
\caption{ The product  $F(Q^2/m_\mu^2) 
 \left[ C_\Pi(Q^2,m_f) - C_\Pi(0,m_f) \right]$ for the case 
 $\beta=6.0$, $\kappa = 0.1345$. The shaded region shows the
 uncertainty. 
  \label{integ}}
\end{center}
 \vspace*{0.5cm}
\end{figure}

  To calculate the value $a_\mu^{had}$ we evaluate the integral
 \begin{equation}
 I^f(m_f) =  \int_0^\infty \frac{d Q^2}{Q^2} 
 F\left(\frac{Q^2}{m_\mu^2}\right) \left[
 C_\Pi(Q^2,m_f) - C_\Pi(0,m_f) \right]
 \label{Idef}
 \end{equation} 
 for each of our data sets. We then need to extrapolate (or interpolate)
 to the chiral limit (for the $u$ and $d$ quarks) and to the strange 
 quark mass. As can be seen in Fig.{\ref{integ}} the integral is 
 dominated by $Q^2 \sim 3 m_\mu^2$, so we need to extrapolate in
 $Q^2$, as our lowest measured value is at $Q^2 = 0.17 \gev^2$. 
 We do our extrapolation of $C_\Pi$ by using the results of Fit~II 
 from the previous section. 

\begin{figure}[htb] 
\begin{center}
\epsfig{file=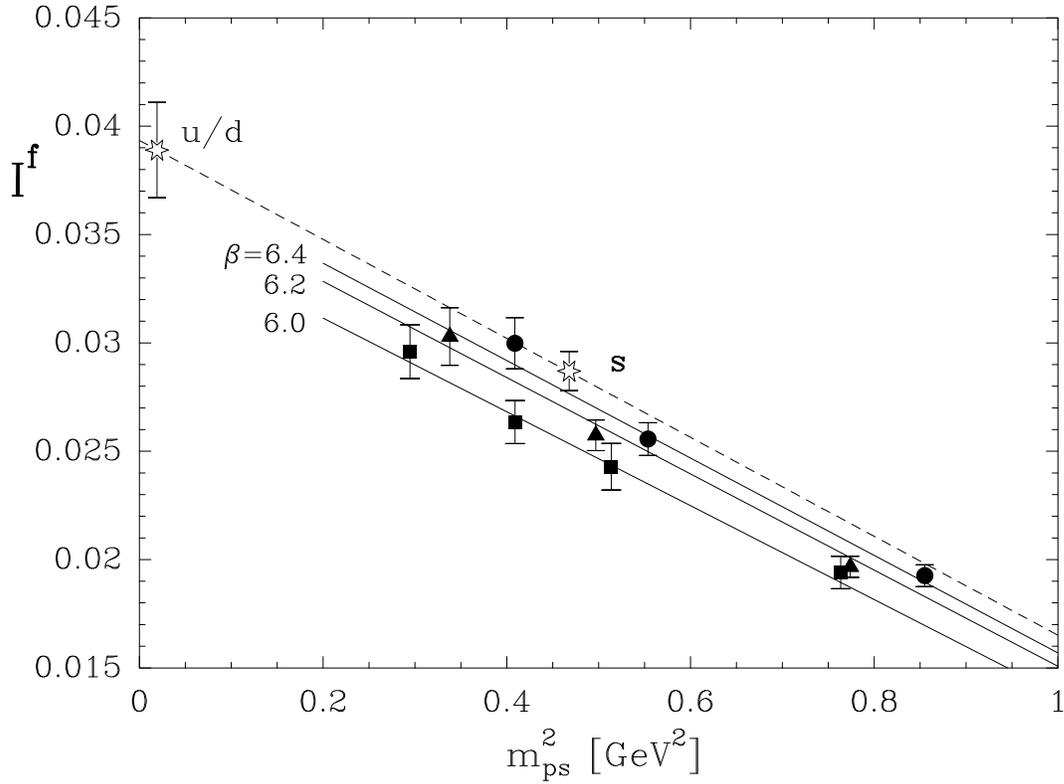,angle=270,width=14cm}
\caption{ The results of the integral~(\ref{Idef}). 
 The $\beta$ values are shown by the same symbols as in Fig.\ref{s0pic}. 
 The dashed line shows our extrapolation to the continuum limit, 
 and the points shown with stars are our results for the light ($u,d$)
 and strange quarks. 
  \label{Ivalfig}}
\end{center}
 \vspace*{0.5cm}
\end{figure}

   The results are shown in Fig.\ref{Ivalfig},
 plotted against $m_{PS}^2$ (the square of the pseudoscalar mass).
 We see that $I$ 
 depends strongly on the quark mass, with heavier quarks making a 
 smaller contribution (as one would expect). There is also some
 dependence on $\beta$. To extrapolate to the continuum we
 fit the data with an ansatz of the form
 \begin{equation}
 I^f = (A_1 + A_2 a^2) + (B_1 + B_2 a^2) m_\pi^2
 \end{equation} 
 This describes the data well ($\chi^2/$dof = .53), and gives
 the continuum limit shown by the dashed line in Fig.\ref{Ivalfig}.
 The physically relevant values of $I^f$  are at the physical pion
 mass ($M^2 = 0.019\gev^2$) for the $u$ and $d$ quarks, and at the
 mass of a hypothetical $\bar{s}s$ pseudoscalar meson with 
 $M^2 = 2 m_K^2 - m_\pi^2 = 0.468\gev^2$ for $I^s$. 
 The values at these points are
 \begin{equation}
  I^u = I^d = 0.0389(21)\,, \quad  I^s  = 0.0287(9) \;.
 \end{equation} 

 This gives as our final lattice estimate of $a_\mu^{had}$ 
 \begin{equation}
 a_\mu^{had} = \frac{\alpha_{em}^2}{3 \pi^2} 
 \left(\frac{4}{9} I^u + \frac{1}{9} I^d + \frac{1}{9} I^s \right) 
 = 446(23) \times 10^{-10}.
 \end{equation} 
 This error reflects the statistical errors of the lattice
 calculation and the extrapolations to the physical points. 
 It does not include any estimate of the error due to the quenched
 approximation used in the calculation. 
 This value is somewhat lower than the experimental value
 $692.4\pm 5.9_{exp}\pm 2.4_{radiative}\times 10^{-10}$,
 found by applying eq.~(\ref{Rint})
 to experimental $R$ measurements~\cite{Eidel2,Rpapers}.
 Our value is similar to another lattice measurement~\cite{Blum}, 
 which finds $460(78) \times 10^{-10}$. 

  The shortfall in the value of $a_\mu^{had}$ can probably be
 attributed to quenching. In particular a quenched calculation 
 omits the process $e^+ e^- \to \pi\pi$, which is the only
 contribution allowed at very low $s$.

 \section{The applicability of the Operator Product Expansion} 

    What does the success of our fit function eq.~(\ref{model})
 tell us about the applicability and usefulness of the OPE? 
 
   To answer this question 
 let us look at the large $Q^2$ behaviour of the formula~(\ref{disp_Ansatz}): 
  \begin{eqnarray}
  C_\Pi(Q^2) &=& B \ln (Q^2) + K + \frac{B s_0 -A}{Q^2}
 + \frac{A m_\rho^2 - \frac{B}{2} s_0^2 }{(Q^2)^2} + \cdots \nonumber\\
 &=&  B \ln (Q^2) + K + \sum_{n=1}^\infty (-1)^n 
  \left\{A (m_\rho^2)^{n-1} -\frac{B}{n} s_0^n\right\} (Q^2)^{-n} . 
 \label{expand} 
 \end{eqnarray}
 The expansion has the same form as the OPE,  at least at leading order
 when there are no logarithmic corrections to the Wilson coefficients.
 We can now look at the higher twist terms in the expansion, and 
 estimate how important they are in comparison with the gluon condensate
 contribution. 

\begin{figure}[htb]
 \vspace*{5mm}
  \epsfig{file=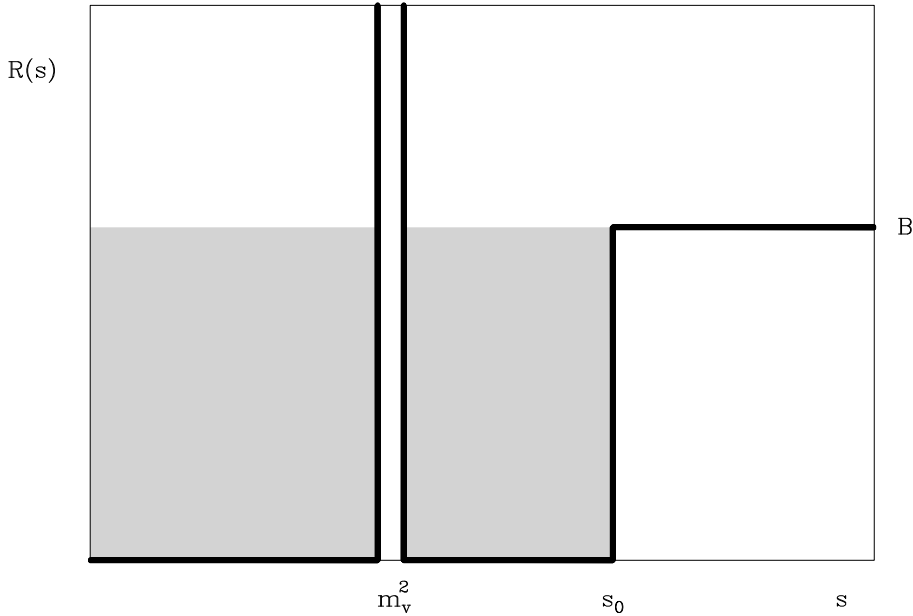,angle=270,width=12cm}
  \caption{
  Our simple model for the contribution of the $u$ and $d$ quarks to 
  $R(s)$. \label{Rmodel} }  
  \end{figure}

 By looking at the first few terms we can relate our parameters 
 $s_0$ and $A$ to the condensates in the OPE. 
 In the chiral limit there is no $1/Q^2$ term, so 
 \begin{equation} 
 A = B s_0 . 
 \end{equation} 
 This says that the area under the meson $\delta$-function is the same
 as the grey area in Fig.\ref{Rmodel}, a typical sum-rule style result. 
  Substituting this into the $1/Q^4$ term we get 
 \begin{equation} 
 B s_0 \left( m_\rho^2 - \frac{1}{2}s_0 \right) = 
 -\pi^2 \frac{\alpha_s}{\pi}\langle G G \rangle  \;.
 \end{equation} 
 Again, a typical sum-rule result --- if there were no gluon 
 condensate, the meson would lie at $\frac{1}{2}s_0$, exactly 
 in the middle of the gap. Putting $m_\rho^2 = 0.6\gev^2$ 
 (the physical value)
 and $\frac{\alpha_s}{\pi}\langle G G \rangle  = 0.012\gev^4$ 
 gives $s_0 = 1.26 \gev^2$. 

\begin{figure}[htb] 
 \epsfig{file=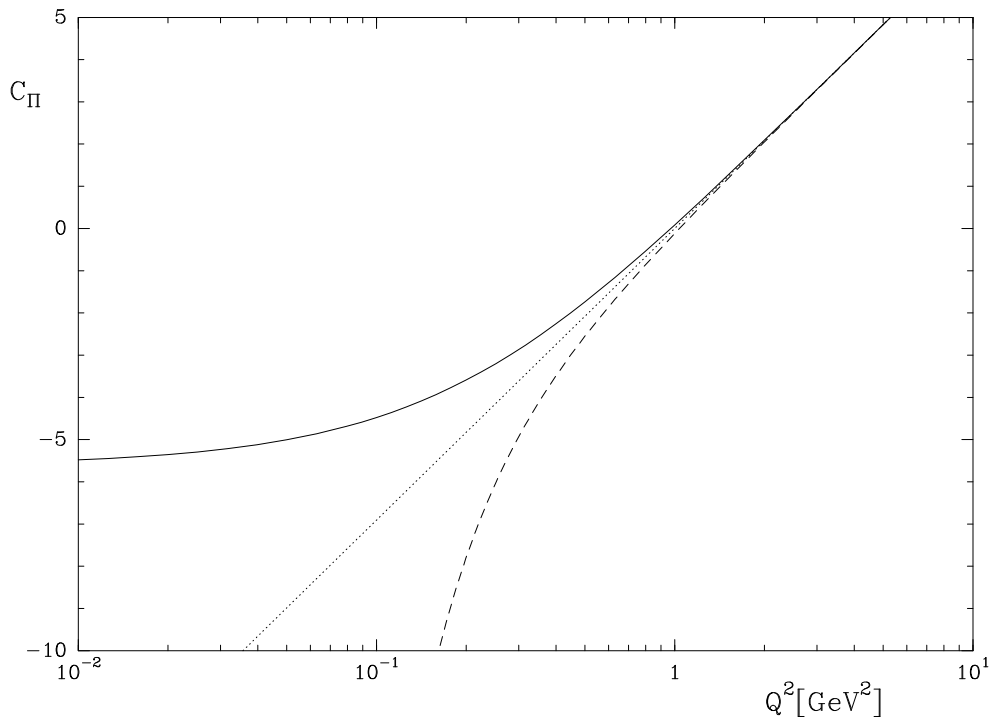,angle=270,width=13cm}
 \caption{  A comparison of the threshold model, eq.(\ref{disp_Ansatz})
 (solid line), perturbation theory (dotted line) and 
 perturbation theory plus the gluon condensate contribution
 (dashed line). In this plot we have chosen the physically
 irrelevant constant $K$ to be 0 when all quantities
 are expressed in GeV. 
 \label{unsub}}
 \end{figure}

 Next, in Fig.\ref{unsub},
 we plot a comparison of the threshold model, eq.(\ref{disp_Ansatz}), 
 perturbation theory (just $3 \ln a^2 Q^2$ at this level) and 
 perturbation theory plus the gluon condensate contribution,
 which scales like $1/(Q^2)^2$.  
 The physically irrelevant constant $K$ has been set to 0 in all cases.  

   The curve from the threshold model looks physically
 sensible, it goes to a finite value at $Q^2 = 0$, which is
 what must happen if $R(s)$ doesn't extend all the way down to 
 $s=0$. 
 
  However, even though we chose our parameters  $A$ and $s_0$
 so that the threshold
 model would match the gluon condensate prediction at high $Q^2$, 
 the two curves don't resemble each other closely. 

\begin{figure}
 \epsfig{file=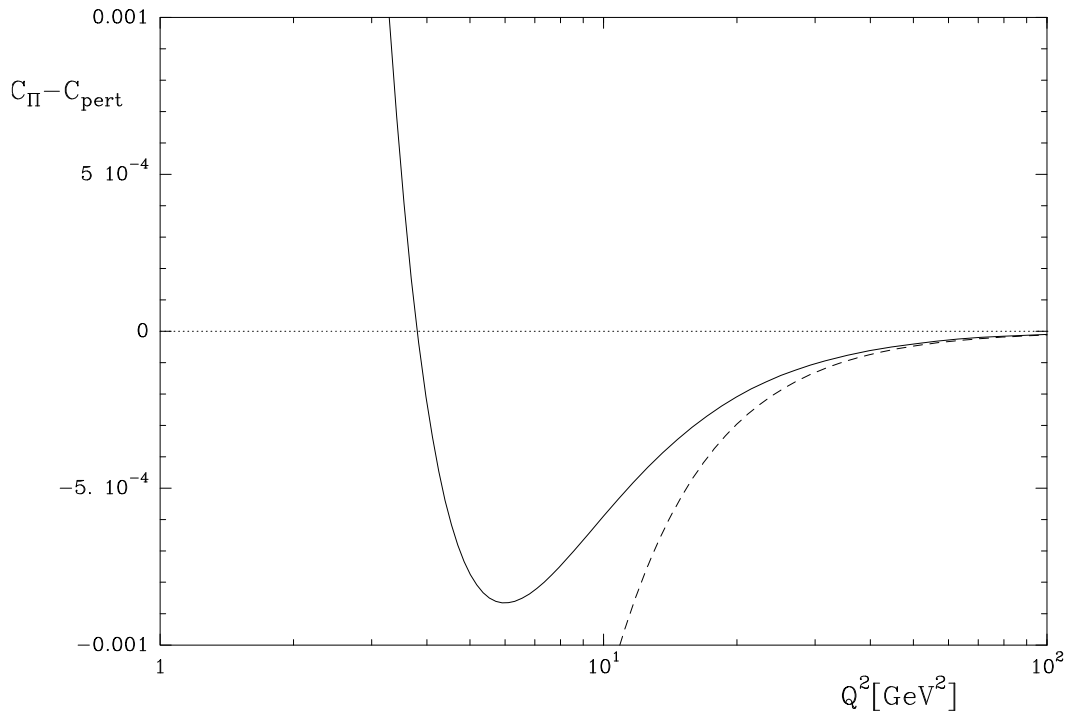,angle=270,width=14cm}
 \caption{ The same as Fig.\ref{unsub}, but with the perturbative
 piece subtracted. Note that in this figure we are concentrating
 on the behaviour at rather large $Q^2$ where non-perturbative
 effects are small. \label{sub} }  
 \end{figure}

    Let us now subtract out the perturbative piece, to see things more
 clearly, Fig.\ref{sub}. We can  only find a region where the gluon
 condensate region is important when we concentrate on
 the large $Q^2$ region. 
    If  Fig.\ref{sub} is what the real world looks like, 
 the prospects for getting at the gluon condensate look poor. 
 The gluon condensate term is overwhelmed by higher order terms
 in the OPE before it has a chance to get significant. 

    What is the conclusion of this exercise? 
  In the region where we can easily see deviations from 
 perturbation theory the OPE is not very useful, because many
 operators of very high dimension are all contributing, 
 not just the leading $1/(Q^2)^2$ contribution coming from the
 gluon condensate. 
 To see the gluon condensate uncontaminated by higher order 
 operators we would need to look in the region 
 $Q^2 \sim 10\gev^2$ with very accurate data (error bars 
 at least 2 orders of magnitude smaller than in this paper).
 This is unfortunately not a realistic prospect. 

\section{Conclusions}

We have computed the vacuum polarisation in the limit of two light flavours
in quenched QCD for three $\beta$ values ($6.0$, $6.2$ and $6.4$) and on 
lattices as large as $32^4$. It was important to improve the action and the 
vector current. We found good agreement with three-loop perturbation theory 
in the interval $2 \lesssim Q^2 \lesssim 20 \,\mbox{GeV}^2$. The 
lattice data show some indication of non-perturbative
 effects at the lower end of the $Q^2$ range. We can describe these
 very well with a model of $R(s)$ which includes vector mesons and 
 threshold effects.  In order to make firm 
predictions, we need more data at small $Q^2$ and at smaller lattice 
spacings with high statistics. 

 From the low $Q^2$ region of the vacuum polarisation  we can
 extract a lattice value for $a_\mu^{had}$, the hadronic
 contribution to the muon's anomalous magnetic moment. 
 We find the value $446(23) \times 10^{-10}$ which is of
 the right order of magnitude, though lower than the physical
 value.  Our estimate could be improved
 by using a larger lattice size, enabling us to reach lower $Q^2$
 which would reduce uncertainties from extrapolation.  Naturally,
  dynamical calculations would be very interesting.

\section*{Acknowledgements}

The numerical calculations were performed on the
Quadrics computers at DESY Zeuthen. We thank the operating staff for their 
support. This work was supported in part by the European Community's Human 
Potential Program under Contract HPRN-CT-2000-00145, Hadrons/Lattice QCD, as 
well as by the DFG (Forschergruppe Gitter-Hadronen-Ph\"anomenologie)
and BMBF.

   We would like to thank C. Michael and T. Teubner for useful comments. 

\section*{Appendix A}

 We denote the quark propagator from lattice points $x$ to $y$ by $G(x,y)$.
 For $\Pi_{\mu \nu}^{(1)}(\hat{q})$ we then obtain
 \begin{equation}
 \begin{split}
 \Pi_{\mu \nu}^{(1)}(\hat{q}) &= \frac{a^4}{4}\sum_{x} e ^{iq(x+
 a\hat{\mu}/2 -a\hat{\nu}/2)} \\
 &\hspace*{0.6cm}\mbox{Tr}
 {\big \langle}(1+\gamma_{\nu})U_{\nu}^{\dagger}(0)\gamma_5 G^{\dagger} 
 (x+a\hat{\mu},0)\gamma_5(1+ \gamma_{\mu})U_{\mu}^{\dagger}(x) G(x,a\hat{\nu})\\
 &\hspace*{0.75cm}- (1-\gamma_{\nu}) U_{\nu}(0)\gamma_5 G^{\dagger}(x 
 +a\hat{\mu},a\hat{\nu}) 
 \gamma_5 (1+ \gamma_{\mu}) U_{\mu}^{\dagger}(x) G(x,0) \\
 &\hspace*{0.75cm}- (1+\gamma_{\nu}) U_{\nu}^{\dagger}(0)\gamma_5 
 G^{\dagger}(x,0)\gamma_5
 (1- \gamma_{\mu})U_{\mu}(x) G(x+a\hat{\mu},a\hat{\nu}) \\
 &\hspace*{0.75cm}+ (1-\gamma_{\nu})U_{\nu}(0)\gamma_5 
 G^{\dagger}(x,a\hat{\nu})\gamma_5
 (1- \gamma_{\mu}) U_{\mu}(x) G(x+a\hat{\mu},0) {\big \rangle} \\
 &-\frac{a^5}{4} \ccvc \sum_{x}  e ^{iq(x+a\hat{\mu}/2)} 
 \hat{q}_\lambda \\
 &\hspace*{0.6cm}\mbox{Tr} {\big \langle}
 (1+ \gamma_{\mu})U_{\mu}^{\dagger}(x) G(x,0) \sigma_{\nu\lambda}\gamma_5
 G^{\dagger}(x+a\hat{\mu},0)\gamma_5\\
 &\hspace*{0.75cm}- (1-\gamma_{\mu}) U_{\mu}(x) G(x+a\hat{\mu},0)
 \sigma_{\nu\lambda}\gamma_5 G^{\dagger}(x,0)\gamma_5 {\big \rangle} \\
 &+\frac{a^5}{4} \ccvc \sum_{x}  e ^{iq(x-a\hat{\nu}/2)} 
 \hat{q}_\sigma \\
 &\hspace*{0.6cm}\mbox{Tr} {\big \langle}
 (1+ \gamma_{\nu}) U_{\nu}^{\dagger}(0)\gamma_5 G^{\dagger}(x,0)\gamma_5
 \sigma_{\mu\sigma} G(x,a\hat{\nu}) \\
 &\hspace*{0.75cm}- (1- \gamma_{\nu}) U_{\nu}(0) \gamma_5 
 G^{\dagger}(x,a\hat{\nu})\gamma_5 \sigma_{\mu \sigma} G(x,0) {\big \rangle} \\
 &-\frac{a^6}{4} \ccvc^2 \sum_{x}  e ^{iqx} 
 \hat{q}_\lambda \hat{q}_\sigma\\
 &\hspace*{0.6cm}\mbox{Tr} {\big \langle}
 \sigma_{\mu \sigma} G(x,0) \sigma_{\nu \lambda} \gamma_5 G^{\dagger}(x,0)
 \gamma_5 {\big \rangle} \, ,
 \end{split}
 \end{equation}
 where the improvement term has been integrated by parts.
 For $\Pi_{\mu \nu}^{(2)}(\hat{q})$ we obtain
 \begin{equation}
 %\begin{split}
 \Pi_{\mu \nu}^{(2)}(\hat{q}) = \frac{a}{2} \delta_{\mu\nu} 
 %\\
 \mbox{Tr} {\big \langle} (1+\gamma_\nu) U^{\dagger}_{\nu}(0)
 G(0,a\hat{\nu}) + (1-\gamma_\nu) U_{\nu}(0) G^{\dagger}(0,a\hat{\nu})
 {\big \rangle} \, .
 %\end{split}
 \end{equation}
 To compute $\Pi_{\mu \nu}^{(1)}(\hat{q})$ and $\Pi_{\mu \nu}^{(2)}(\hat{q})$
 we have to do a minimum of five inversions for each gauge field configuration
 (and each $\kappa$ value),
 which makes the calculation computationally quite expensive, in particular on
 $32^4$ lattices.

\section*{Appendix B}

 \subsection*{Perturbative results}
 
Before we describe the lattice calculation in detail, we present here the 
perturbative Wilson 
coefficients $c_0$, $c_4^F$ and $c_4^G$. We will work in the quenched
approximation, in which contributions from sea quarks are neglected, 
and which corresponds to $c_4^{\,\prime\,F} = 0$.

We write 
 \begin{eqnarray}
\label{ccc} \lefteqn{
c_0(\mu^2,Q^2,m) = c_0^{(0)}(\mu^2,Q^2,m) 
+ \frac{\alpha_s(\mu^2)}{\pi} C_F\, c_0^{(1)}(\mu^2,Q^2,m) }
 \\
&&+ \big(\frac{\alpha_s(\mu^2)}{\pi}\big)^2
 \big(C_F^2\, c_0^{(2)}(\mu^2,Q^2,m)
+ C_F C_A\, c_0^{(2)\,\prime}(\mu^2,Q^2,m)\big)+\cdots,
 \nonumber   
 \end{eqnarray}
with $C_F=4/3$ and $C_A=3$ for $SU(3)$. 
These coefficients can be found in eqs.~(27)-(30) of~\cite{Chetyrkin&Kuhn}  
 (recall that  $Q^2 \equiv -q^2$). 
In the $\msbar$ scheme the coefficients 
$c_0^{(0)}$, $c_0^{(1)}$, $c_0^{(2)}$ and $c_0^{(2)\,\prime}$ 
read 
\begin{equation}
\label{ggz1}
\begin{split}
c_0^{(0)}(\mu^2,Q^2,\bar{m}) = - \frac{9}{4} 
\Bigg[&
\frac{20}{9} -\frac{4}{3}\ln\frac{Q^2}{\mu^2}
-8\frac{\bar{m}^2}{Q^2}
+\big(\frac{4\bar{m}^2}{Q^2}\big)^2
\big(\frac{1}{4}+\frac{1}{2}\ln\frac{Q^2}{\bar{m}^2}\big)\Bigg]
\, , %\\[0.5em]
\end{split}
\end{equation}
\begin{equation}
\label{ggz2}
\begin{split}
c_0^{(1)}(\mu^2,Q^2,\bar{m}) = - \frac{9}{4} \Bigg[&
     \frac{55}{12} - 4\zeta(3)  - \ln\frac{Q^2}{\mu^2}
   -\frac{4\bar{m}^2}{Q^2}\big(
          4 - 3\ln\frac{Q^2}{\mu^2}
                    \big) \\
&+\big(\frac{4\bar{m}^2}{Q^2}\big)^2 \big(
       \frac{1}{24} + \zeta(3)
      +\frac{11}{8}\ln\frac{Q^2}{\bar{m}^2}
 \\ &
      +\frac{3}{4}\ln^2\frac{Q^2}{\bar{m}^2} 
      -\frac{3}{2}\ln\frac{Q^2}{\bar{m}^2}\ln\frac{Q^2}{\mu^2}
\big) \Bigg] \, , %\\[0.5em]
\end{split}
\end{equation}
\begin{equation}
\label{ggz3}
\begin{split}
c_0^{(2)}(\mu^2,Q^2,\bar{m}) = - \frac{9}{4}  \Bigg[&
       -\frac{143}{72} 
       - \frac{37}{6}\zeta(3)
       + 10\zeta(5)
       + \frac{1}{8}\ln\frac{Q^2}{\mu^2} \\
&-\frac{4\bar{m}^2}{Q^2}\big(
       \frac{1667}{96} 
     - \frac{5}{12}\zeta(3) 
- \frac{35}{6}\zeta(5)
 \\ &
     - \frac{51}{8}\ln\frac{Q^2}{\mu^2} 
     + \frac{9}{4}\ln^2\frac{Q^2}{\mu^2}
\big)\Bigg] \, , %\\[0.5em]
\end{split}
\end{equation}
\begin{equation}
\label{ggz4}
\begin{split}
c_0^{(2)\,\prime}(\mu^2,Q^2,\bar{m}) = - \frac{9}{4}
\Bigg[& \frac{44215}{2592} 
       - \frac{227}{18}\zeta(3)
       - \frac{5}{3}\zeta(5)
       - \frac{41}{8}\ln\frac{Q^2}{\mu^2} \\
&       + \frac{11}{24}\ln^2\frac{Q^2}{\mu^2}
+ \frac{11}{3}\zeta(3)\ln\frac{Q^2}{\mu^2}
 \\ &
-\frac{4\bar{m}^2}{Q^2}\big(
         \frac{1447}{96} 
       + \frac{4}{3}\zeta(3)
       - \frac{85}{12}\zeta(5)
 \\ &
       - \frac{185}{24}\ln\frac{Q^2}{\mu^2} 
 + \frac{11}{8}\ln^2\frac{Q^2}{\mu^2}
\big)\Bigg] \, ,
\end{split}
\end{equation}
where $\zeta(3)=1.20206\cdots$ and $\zeta(5)=1.03693\cdots$, and $\bar{m}$
refers to the quark mass at the scale $\mu$
 in the $\msbar$ scheme. In (\ref{ggz1}),(\ref{ggz2})
terms of $\mathcal{O}((\bar{m}^2/Q^2)^3)$ modulo logarithms
 have been neglected,
 while in (\ref{ggz3}),(\ref{ggz4}) terms 
  $\mathcal{O}((\bar{m}^2/Q^2)^2)$ are dropped. 

The Wilson coefficients multiplying the quark and gluon condensate 
are~\cite{SVZI,Wilson_NP}
\begin{equation}
\begin{split}
c_4^F(\mu,q) &= -12 \pi^2
\Big(2 + \frac{2}{3}
\frac{\alpha_s(\mu^2)}{\pi}
 +\cdots \Big) \, , \\[0.5em]
c_4^G(\mu,q) &= -\pi^2 \big(1-\frac{11}{18}\frac{\alpha_s(\mu^2)}{\pi}
+\cdots \big) 
\, .
\end{split}
\end{equation}
Note that both $m \langle \bar{q} q\rangle$ and $(\alpha_s/\pi)\, 
\langle G_{\mu\nu}^2\rangle$ are renormalisation group invariants, which means
that they do not depend on $\mu$ and the renormalisation scheme. The vacuum 
polarisation $\Pi(Q^2)$ itself is not an observable, but its derivatives are. 
Therefore the result (\ref{OPE}) can only depend on $\mu$ and the scheme in
terms of an integration constant (independent of $Q^2$).

\subsection*{Renormalisation Group Improvement} 

  We can use renormalisation group improvement to re-sum  the
 logarithms in higher-order terms. This should lead to a significant
 improvement, since the fact that we are interested in measurements over
 a large $Q^2$ range means that these logarithms are large. 
 
  If we calculate the Adler function in the chiral limit
 from eqs~(\ref{ggz1})-(\ref{ggz4}) we find the result
 \begin{equation} 
 D(Q^2) = 3 \Bigg\{ 1 + \frac{\alpha_s(\mu^2)}{\pi}
 + \left( \frac{\alpha_s(\mu^2)}{\pi} \right)^2 
 \left[ \frac{365}{24} - 11 \zeta(3) - \frac{11}{4} \ln
 \frac{Q^2}{\mu^2} \right] \Bigg\} . 
 \end{equation} 
 We should be able to do better than this because the 
 massless Adler function is known to four loops~\cite{Gorish}. 
  We can use this perturbative result for the Adler
 function to improve the result for $c_0$ in the
 chiral limit. 
 In quenched $SU(3)$ we have 
  \begin{eqnarray} 
 \label{Rpert}
  R(s)  =& 3 \sum_f e_f^2 & \Bigg\{ 1 + \asQ  
  + \asQ^2 \left[ \frac{365}{24} - 11 \zeta(3) \right] 
  \nonumber \\ && 
  + \asQ^3 \left[ \frac{87029}{288} -\frac{1103}{4} \zeta(3) 
  + \frac{275}{6} \zeta(5) -\frac{121}{48} \pi^2 \right] 
   \Bigg\}
  \nonumber \\ &  
  + ( \sum_f e_f )^2 &
   \asQ^3 \left[ \frac{55}{72} - \frac{5}{3} \zeta(3) \right] 
  + \mathcal{O}(\alpha_s^4) 
  \end{eqnarray} 
  and 
 \begin{eqnarray} 
 \label{Dpert} 
 D(Q^2)  =& 3 \sum_f e_f^2 &\Bigg\{ 1 + \asQ  
 + \asQ^2 \left[ \frac{365}{24} - 11 \zeta(3) \right] 
 \nonumber \\ && 
 + \asQ^3 \left[ \frac{87029}{288} -\frac{1103}{4} \zeta(3) 
 + \frac{275}{6} \zeta(5) \right] \Bigg\}
 \nonumber \\ &  
 + ( \sum_f e_f )^2& 
  \asQ^3 \left[ \frac{55}{72} - \frac{5}{3} \zeta(3) \right] 
 + \mathcal{O}(\alpha_s^4) \; .
 \end{eqnarray} 
% where $Q^2 \equiv -q^2$. 
  Note that although the first terms of $R$ and $D$ coincide, 
  the $\alpha_s^3$ term is different. The extra "$\pi^2$'' term 
 in $R$ arises
  from analytic continuation, 
 $\ln^3(Q^2/\mu^2) \rightarrow [\ln(s/\mu^2) \pm i \pi]^3 $,
 see~\cite{Gorish}. The first part of (\ref{Dpert}), 
 proportional to $\sum_f e_f^2$, is the derivative of the $C_\Pi$
 term in (\ref{pisplit}), while the second term, proportional to 
 $( \sum_f e_f )^2$, comes from the derivative of the $A_\Pi$ term.

 From (\ref{Dpert}) we have a differential equation for $c_0$: 
 \begin{eqnarray}
 \lefteqn{ \!\!\!
   Q^2 \frac{\partial}{\partial Q^2} c_0(\mu^2, Q^2,m\!=0)  
   = 3 \Bigg\{ 1 + \asQ
  + \asQ^2 \left[ \frac{365}{24} - 11 \zeta(3) \right]}
 \nonumber \\ && \hspace{1.2cm}{}   
 + \asQ^3 \left[ \frac{87029}{288} -\frac{1103}{4} \zeta(3)
 + \frac{275}{6} \zeta(5) \right] +\mathcal{O}(\alpha_s^4) \Bigg\}. 
 \end{eqnarray} 
 We can solve this by using the known $\beta$-function~\cite{betaMS} 
 \begin{equation} 
   Q^2 \frac{\partial}{\partial Q^2} \asQ 
 = - \sum_i \beta_i \asQ^{i+2}
 \end{equation} 
 with 
 \begin{equation} 
 \beta_0 = \frac{11}{4}, \hspace*{8mm}
 \beta_1 = \frac{51}{8},  \hspace*{8mm}
 \beta_2 = \frac{2857}{128}, \hspace*{8mm}
 \beta_3 = \frac{149753}{1536} + \frac{891}{64} \zeta(3)  
 \end{equation}
 in the case of quenched $SU(3)$.  

   The solution is 
 \begin{eqnarray} 
 \lefteqn{ 
 c_0(\mu^2, Q^2,0)  =  3 \ln Q^2 
  -\frac{12}{11} \ln (\alpha_s(Q^2) ) 
 +\left[ -\;\frac{3403}{242} + 12 \zeta(3) \right]
  \frac{\alpha_s(Q^2)}{\pi} } 
  \\ &&     
 +\left[ -\;\frac{2301587}{15972} 
 +\frac{273 }{2} \zeta(3) -25 \zeta(5) \right] 
 \left( \frac{\alpha_s(Q^2)}{\pi} \right)^2 
 + \mathcal{O}(\alpha_s^3) 
 + {\rm const} . 
 \nonumber 
 \end{eqnarray} 
 The constant of integration can be fixed by comparing with 
 (\ref{ccc})-(\ref{ggz4}), giving the final renormalisation
 group improved result\footnote{Note the interesting result
 that due to a cancellation there is no $\zeta(3)$ term
 in the coefficient of $\alpha_s(\mu^2)$ and no $\zeta(5)$ in
 the coefficient of  $\alpha_s(\mu^2)^2$. This surprised us.} 
 \begin{eqnarray} 
 \label{improved}  
 c_0(\mu^2, Q^2,0)  &=&  3 \ln \frac{Q^2}{\mu^2} 
  -\frac{12}{11} \ln \left[\frac{\alpha_s(Q^2)}{\alpha_s(\mu^2)} \right] 
 +\left[ -\;\frac{3403}{242} + 12 \zeta(3) \right]
  \frac{\alpha_s(Q^2)}{\pi}   
   \nonumber \\ &&     
 +\left[ -\;\frac{2301587}{15972} 
 +\frac{273 }{2} \zeta(3) -25 \zeta(5) \right] 
 \left( \frac{\alpha_s(Q^2)}{\pi} \right)^2 
   \\ && 
  -5 + \frac{151}{484} \,\frac{\alpha_s(\mu^2)}{\pi}  
  +\left[  -\;\frac{566749}{383328} + \frac{5}{3} \zeta(3) \right] 
  \left( \frac{\alpha_s(\mu^2)}{\pi}\right)^2 
  \nonumber \\ && 
  + O\left(\alpha_s^3(Q^2),\alpha_s^3(\mu^2)\right)   
 \nonumber  
 \end{eqnarray} 
 where $\mu$ is the $\msb$ scale. 
 
  One can check that the differences between (\ref{improved})
 and (\ref{ccc}) are indeed $ O(\alpha_s^3)$ by 
 making the substitution 
 \begin{eqnarray} 
 \frac{\alpha_s(Q^2)}{\pi} &\rightarrow& \frac{\alpha_s(\mu^2)}{\pi} 
 -\beta_0 \left(\frac{\alpha_s(\mu^2)}{\pi}\right)^2
  \ln \frac{Q^2}{\mu^2} 
  -\beta_1 \left(\frac{\alpha_s(\mu^2)}{\pi}\right)^3
  \ln \frac{Q^2}{\mu^2} 
 \nonumber \\ && 
 + \beta_0^2 \left(\frac{\alpha_s(\mu^2)}{\pi}\right)^3
 \ln^2 \frac{Q^2}{\mu^2} + \mathcal{O}(\alpha_s^4)
 \end{eqnarray} 
 in (\ref{improved}). 

    Renormalisation group 
 improvement of the mass terms in (\ref{ggz1})-(\ref{ggz4}) 
 is simpler because there is no constant of integration involved. 
 The result is 
 \begin{eqnarray} \label{c0massive} \lefteqn{
 c_0(\mu^2,Q^2,\bar{m})  =   c_0(\mu^2,Q^2,0) 
 + \frac{m^2_{\msb}(Q^2)}{Q^2} 
 \left\{ 18 + 48 \frac{\alpha_s(Q^2)}{\pi} \right. }
  \nonumber \\  && \left. 
 + \left[ \frac{19691}{24} + \frac{124}{3} \zeta(3) 
 - \frac{1045}{3} \zeta(5) \right] 
  \left(\frac{\alpha_s(Q^2)}{\pi}\right)^2 \right\} \\
 && + \left(\frac{m^2_{\msb}(Q^2)}{Q^2}\right)^2 \left\{ 
 -9 - 18 \ln \frac{Q^2}{m^2_{\msb}(Q^2)} \right.
 \nonumber \\
 && \left. - \left[ 2 + 48 \zeta(3) +
  66 \; \ln \frac{Q^2}{m^2_{\msb}(Q^2)}
 +36 \;\ln^2  \frac{Q^2}{m^2_{\msb}(Q^2)} \right] 
  \frac{\alpha_s(Q^2)}{\pi}
  \right\} .
 \nonumber
 \end{eqnarray}

 What have we gained by using the renormalisation group? 
 Most importantly, the $Q^2$ derivative of (\ref{improved}) is correct
 to four loops, one order better than (\ref{ccc}). 
 The uncertainties in (\ref{improved}) are
 $\mathcal{O}(\alpha_s^3(Q^2),\alpha_s^3(\mu^2))$ while in the 
 original formula (\ref{ccc}) terms of order
  $\mathcal{O}(\alpha_s^3(\mu^2) \ln^3(Q^2/\mu^2) )$ have been neglected. 
 So, if we are interested in describing a large range of 
 $Q^2$, eq.(\ref{improved}) ought to be the better formula.  
 In addition, (\ref{improved}) exhibits correct physics, because it
 is the sum of a piece depending 
 only on $Q^2$ and a piece depending only on $\mu^2$, which is
 only approximately the case for (\ref{ccc}). 
 
\section*{Appendix C}

In the following Tables we give the results for $C_{\Pi}(\hat{q},am)$ at the 
various $\beta$ and $\kappa$ values. We label the lattice momenta by the 
vector $n$, $\hat{q}_\mu = (2/a)\sin(\pi n_\mu/L)$. The momenta were chosen 
close to the diagonal of the Brillouin zone
 to avoid large $\mathcal{O}(a^2)$ effects.

\begin{table}[htbp]
\begin{center}
\begin{small}
\begin{tabular}{ |l|l|l|l|l| }
\hline
\multicolumn{1}{|c|}{$n$}& \multicolumn{1}{|c|}{$\kappa=0.1333$} &
\multicolumn{1}{|c|}{$\kappa=0.1339$} & 
\multicolumn{1}{|c|}{$\kappa=0.1342$} & \multicolumn{1}{|c|}{$\kappa=0.1345$}\\
\hline
(1,0,0,0)&-38.15(11) &-38.60(15) &-38.84(14)&-39.03(16) \\            
(1,1,0,0)&-36.59(10) &-36.89(13) &-37.05(12)&-37.17(13) \\            
(1,1,1,0)&-35.52(09) &-35.75(12) &-35.89(11)&-35.99(12) \\            
(1,1,1,1)&-34.72(09) &-34.91(12) &-35.04(11)&-35.12(12) \\            
(2,1,1,1)&-32.89(08) &-33.06(10) &-33.15(09)&-33.24(10) \\            
(2,2,1,1)&-31.82(07) &-31.98(10) &-32.07(09)&-32.15(10) \\            
(2,2,2,1)&-31.07(07) &-31.22(09) &-31.30(08)&-31.40(09) \\            
(2,2,2,2)&-30.47(07) &-30.62(09) &-30.71(08)&-30.81(09) \\            
(3,2,2,2)&-29.61(06) &-29.76(09) &-29.85(08)&-29.95(09) \\            
(3,3,2,2)&-28.98(06) &-29.13(08) &-29.22(08)&-29.32(08) \\            
(3,3,3,2)&-28.47(06) &-28.63(08) &-28.72(07)&-28.82(08) \\            
(3,3,3,3)&-28.05(06) &-28.20(08) &-28.29(07)&-28.39(08) \\            
(4,3,3,3)&-27.54(05) &-27.70(08) &-27.79(07)&-27.89(08) \\            
(4,4,3,3)&-27.12(05) &-27.28(07) &-27.37(07)&-27.47(08) \\            
(4,4,4,3)&-26.76(05) &-26.91(07) &-27.00(07)&-27.11(07) \\            
(4,4,4,4)&-26.42(05) &-26.58(07) &-26.67(07)&-26.78(07) \\            
(5,4,4,4)&-26.11(05) &-26.26(07) &-26.36(06)&-26.46(07) \\            
(5,5,4,4)&-25.82(05) &-25.97(07) &-26.07(06)&-26.17(07) \\            
(5,5,5,4)&-25.55(05) &-25.70(07) &-25.80(06)&-25.90(07) \\            
(5,5,5,5)&-25.29(05) &-25.44(07) &-25.54(06)&-25.64(07) \\            
(6,5,5,5)&-25.09(05) &-25.25(06) &-25.34(06)&-25.45(07) \\     
(6,6,5,5)&-24.90(04) &-25.05(06) &-25.15(06)&-25.25(06) \\            
(6,6,6,5)&-24.71(04) &-24.86(06) &-24.96(06)&-25.06(06) \\            
(6,6,6,6)&-24.52(04) &-24.67(06) &-24.77(06)&-24.87(06) \\            
(7,6,6,6)&-24.42(04) &-24.57(06) &-24.67(05)&-24.77(06) \\            
(7,7,6,6)&-24.31(04) &-24.46(06) &-24.56(05)&-24.66(06) \\            
(7,7,7,6)&-24.19(04) &-24.34(06) &-24.44(05)&-24.54(06) \\            
(7,7,7,7)&-24.07(04) &-24.22(06) &-24.32(05)&-24.42(06) \\ \hline 
\end{tabular}
\end{small}
\end{center}
\vspace*{0.5cm}
\caption{The vacuum polarisation $C_\Pi(\hat{q}^2,am)$ on the $16^4$ 
lattice at $\beta=6.0$.} 
\end{table}

\clearpage 

\begin{table}[htbp]
\begin{center}
\begin{small}
\begin{tabular}{ |l|l| }
\hline
\multicolumn{1}{|c|}{$n$}& \multicolumn{1}{|c|}{$\kappa=0.1345$} \\
\hline
 (1,0,0,0) &   -41.64(22) \\
 (1,1,0,0) &   -40.67(19) \\
 (1,1,1,0) &   -39.85(17) \\
 (1,1,1,1) &   -39.16(16) \\
 (2,1,1,1) &   -37.58(14) \\
 (2,2,1,1) &   -36.51(13) \\
 (2,2,2,1) &   -35.72(13) \\
 (2,2,2,2) &   -35.09(12) \\
 (3,2,2,2) &   -34.20(12) \\
 (3,3,2,2) &   -33.52(11) \\
 (3,3,3,2) &   -32.99(11) \\
 (3,3,3,3) &   -32.54(11) \\
 (4,3,3,3) &   -31.97(11) \\
 (4,4,3,3) &   -31.50(10) \\
 (4,4,4,3) &   -31.11(10) \\
 (4,4,4,4) &   -30.77(10) \\
 (5,4,4,4) &   -30.36(10) \\
 (5,5,4,4) &   -30.01(10) \\
 (5,5,5,4) &   -29.70(10) \\
 (5,5,5,5) &   -29.43( 9) \\
 (6,5,5,5) &   -29.11( 9) \\
 (6,6,5,5) &   -28.84( 9) \\
 (6,6,6,5) &   -28.59( 9) \\
 (6,6,6,6) &   -28.36( 9) \\
 (7,6,6,6) &   -28.11( 9) \\
 (7,7,6,6) &   -27.88( 9) \\
 (7,7,7,6) &   -27.68( 9) \\
 (7,7,7,7) &   -27.48( 9) \\
\hline 
\end{tabular}
\end{small}
\end{center}
\vspace*{0.5cm}
\caption{The vacuum polarisation $C_\Pi(\hat{q}^2,am)$ on the $32^4$
lattice at $\beta=6.0$.}
\end{table}

\begin{table}[htbp]
\begin{center}
\begin{small}
\begin{tabular}{ |l|l|l|l| }
\hline
\multicolumn{1}{|c|}{$n$}& \multicolumn{1}{|c|}{$\kappa=0.1344$} &
\multicolumn{1}{|c|}{$\kappa=0.1349$} & 
\multicolumn{1}{|c|}{$\kappa=0.1352$} \\
\hline
(1,0,0,0)       &-40.24(14)      &-40.87(17)     &-41.16(16)\\
(1,1,0,0)       &-38.78(12)      &-39.17(14)     &-39.35(12)\\
(1,1,1,0)       &-37.74(11)      &-38.03(13)     &-38.17(11)\\
(1,1,1,1)       &-36.96(10)      &-37.19(12)     &-37.30(11)\\
(2,1,1,1)       &-35.23(09)      &-35.40(10)     &-35.47(09)\\
(2,2,1,1)       &-34.16(08)      &-34.31(09)     &-34.38(08)\\
(2,2,2,1)       &-33.39(07)     &-33.54(08)     &-33.60(07)\\
(2,2,2,2) &-32.79(07)    &-32.94(08)    &-33.00(07)\\
(3,2,2,2) &-31.93(06)    &-32.08(07)    &-32.13(07)\\
(3,3,2,2) &-31.29(06)    &-31.44(07)    &-31.49(06)\\
(3,3,3,2) &-30.78(06)    &-30.94(07)    &-30.99(06)\\
(3,3,3,3) &-30.35(06)    &-30.52(06)    &-30.57(06)\\
(4,3,3,3) &-29.81(06)    &-29.98(06)    &-30.03(06)\\
(4,4,3,3) &-29.37(06)    &-29.54(06)    &-29.59(06)\\
(4,4,4,3) &-28.99(05)    &-29.17(06)    &-29.22(06)\\
(4,4,4,4) &-28.67(05)    &-28.84(06)    &-28.90(05)\\
(5,4,4,4) &-28.29(05)    &-28.46(06)    &-28.52(05)\\
(5,5,4,4) &-27.96(05)    &-28.14(05)    &-28.20(05)\\
(5,5,5,4) &-27.67(05)    &-27.85(05)    &-27.91(05)\\
(5,5,5,5) &-27.41(05)    &-27.58(05)    &-27.64(05)\\
(6,5,5,5) &-27.12(05)    &-27.30(05)    &-27.37(05)\\
(6,6,5,5) &-26.87(05)    &-27.05(05)    &-27.11(05)\\
(6,6,6,5) &-26.64(05)    &-26.82(05)    &-26.89(05)\\
(6,6,6,6) &-26.42(05)    &-26.60(05)    &-26.66(05)\\
(7,6,6,6) &-26.21(05)    &-26.39(05)    &-26.45(05)\\
(7,7,6,6) &-26.01(05)    &-26.19(05)    &-26.26(05)\\
(7,7,7,6) &-25.82(04)    &-26.00(05)    &-26.07(04)\\
(7,7,7,7) &-25.64(04)    &-25.82(05)    &-25.89(04)\\\hline
\end{tabular}
\end{small}
\end{center}
\vspace*{0.5cm}
\caption{The vacuum polarisation $C_\Pi(\hat{q}^2,am)$ on the $24^4$ 
lattice at $\beta=6.2$.} 
\end{table}

\begin{table}[htbp]
\begin{center}
\begin{tabular}{ |l|l|l|l| }
\hline
\multicolumn{1}{|c|}{$n$}& \multicolumn{1}{|c|}{$\kappa=0.1346$} &
\multicolumn{1}{|c|}{$\kappa=0.1350$} & 
\multicolumn{1}{|c|}{$\kappa=0.1352$} \\
\hline
(1,0,0,0)       &-41.88(16)      &-42.52(22)  &-42.84(18)\\
(1,1,0,0)       &-40.38(15)      &-40.77(19)  &-40.97(15)\\
(1,1,1,0)       &-39.33(14)      &-39.61(17)  &-39.75(14)\\
(1,1,1,1)       &-38.53(13)      &-38.74(17)  &-38.86(14)\\
(2,1,1,1)       &-36.82(11)      &-36.96(13)  &-37.02(11)\\
(2,2,1,1)       &-35.73(10)      &-35.84(11)  &-35.89(10)\\
(2,2,2,1)       &-34.93(09)      &-35.04(10)  &-35.09(09)\\
(2,2,2,2)       &-34.31(08)      &-34.42(10)  &-34.46(08)\\
(3,2,2,2)       &-33.44(08)      &-33.55(08)  &-33.58(08)\\
(3,3,2,2)       &-32.79(07)      &-32.89(08)  &-32.93(07)\\
(3,3,3,2) &-32.27(07)     &-32.37(07)    &-32.41(07)\\
(3,3,3,3) &-31.83(06)     &-31.94(07)    &-31.98(06)\\
(4,3,3,3) &-31.28(06)     &-31.39(06)    &-31.42(06)\\
(4,4,3,3) &-30.83(06)     &-30.94(06)    &-30.98(06)\\
(4,4,4,3) &-30.45(05)     &-30.56(06)    &-30.60(05)\\
(4,4,4,4) &-30.12(05)     &-30.23(06)    &-30.28(05)\\
(5,4,4,4) &-29.73(05)     &-29.84(05)    &-29.88(05)\\
(5,5,4,4) &-29.39(05)     &-29.51(05)    &-29.55(05)\\
(5,5,5,4) &-29.09(05)     &-29.21(05)    &-29.25(05)\\
(5,5,5,5) &-28.83(04)     &-28.95(05)    &-28.99(05)\\
(6,5,5,5) &-28.53(04)     &-28.65(05)    &-28.69(04)\\
(6,6,5,5) &-28.26(04)     &-28.39(05)    &-28.43(04)\\
(6,6,6,5) &-28.02(04)     &-28.15(04)    &-28.19(04)\\
(6,6,6,6) &-27.80(04)     &-27.93(04)    &-27.97(04)\\
(7,6,6,6) &-27.56(04)     &-27.70(04)    &-27.74(04)\\
(7,7,6,6) &-27.35(04)     &-27.48(04)    &-27.52(04)\\
(7,7,7,6) &-27.15(04)     &-27.28(04)    &-27.32(04)\\
(7,7,7,7) &-26.96(04)     &-27.10(04)    &-27.14(04)\\
\hline
\end{tabular}
\end{center}
\vspace*{0.5cm}
\caption{The vacuum polarisation $C_\Pi(\hat{q}^2,am)$ on the $32^4$ 
lattice at $\beta=6.4$.}
\end{table}

% ----------------------------------------------------------------------

%
\end{document}